\documentclass{article}
\pdfoutput=1

\usepackage[glossaries=false,manualplacement=true]{hep-paper}

\usepackage{xcolor}
\definecolor{nicered}{rgb}{.7,.1,.1}
\definecolor{nicegreen}{rgb}{.1,.5,.1}
\definecolor{darkblue}{rgb}{0,0,.5}
\hypersetup{colorlinks, citecolor=nicegreen, linkcolor=nicered, urlcolor=darkblue}

\DeclareFieldFormat{pages}{\mkfirstpage[{\mkpageprefix[bookpagination]}]{#1}} 

\graphics{figs}

\newcommand{\epem}{e^+e^-}
\newcommand{\nb}{nb$^{-1}$\xspace}
\newcommand{\sqrts}{\sqrt{s}}
\newcommand{\gaga}{\gamma\gamma}
\newcommand{\bbbar}{b\bar{b}}
\newcommand{\sqrtsnn}{\sqrt{s_{_\text{NN}}}}
\newcommand{\pT}{\ensuremath{p_\mathrm{T}}}

\bibliography{bibliography}

\title{Opportunities for new physics searches with heavy ions at colliders}

\author[cern-ep]{David~d'Enterria}
\author[louvain]{Marco~Drewes}
\author[louvain]{Andrea~Giammanco}
\author[lisboa]{Jan~Hajer}
\author[GSI,GU]{Elena~Bratkovskaya}
\author[cern-c]{Roderik~Bruce}
\author[pnpi,mipt]{Nazar~Burmasov}
\author[agh]{Mateusz~Dyndal}
\author[notts]{Oliver~Gould}
\author[agh]{Iwona~Grabowska-Bold}
\author[GSI]{Malgorzata~Gumberidze}
\author[tokyo]{Taku~Gunji}
\author[GSI]{Romain~Holzmann}
\author[cern-c]{John~M.~Jowett}
\author[pnpi]{Evgeny~Kryshen}
\author[mephi]{Vitalii~A.~Okorokov}
\author[GU]{Ida~Schmidt}
\author[alab]{Aditya~Upreti}

\affiliation[cern-ep]{CERN, EP Department, 1211 Geneva, Switzerland}
\affiliation[louvain]{Centre for Cosmology, Particle Physics and Phenomenology (CP3), Universit\'{e} catholique de Louvain, 1348 Louvain-la-Neuve, Belgium}
\affiliation[lisboa]{Centro de F\'isica Te\'orica de Part\'iculas, Instituto Superior T\'ecnico, Universidade de Lisboa, Portugal}
\affiliation[GSI]{GSI Helmholtzzentrum f\"ur Schwerionenforschung GmbH, 64291 Darmstadt, Germany}
\affiliation[GU]{Institut f\"ur Theoretische Physik, J.~W.~Goethe-Universit\"at, 60438 Frankfurt am Main, Germany}
\affiliation[cern-c]{CERN, BE Department, 1211 Geneva, Switzerland}
\affiliation[pnpi]{Petersburg Nuclear Physics Institute of NRC (Kurchatov Institute), 188300 Gatchina, Russia}
\affiliation[mipt]{Moscow Institute for Physics and Technology, Moscow, Russia}
\affiliation[agh]{AGH University of Science and Technology, Krakow, Poland}
\affiliation[notts]{University of Nottingham, Nottingham NG7 2RD, UK}
\affiliation[tokyo]{Center for Nuclear Study, the Graduate School of Science, University of Tokyo, Tokyo, Japan}
\affiliation[mephi]{National Research Nuclear University MEPhI, 115409 Moscow, Russia}
\affiliation[alab]{The University of Alabama, Tuscaloosa, Alabama, USA}

\begin{document}

\maketitle

\begin{abstract}
Opportunities for searches for phenomena beyond the Standard Model (BSM) using heavy-ions beams at high energies are outlined. Different BSM searches proposed in the last years in collisions of heavy ions, mostly at the Large Hadron Collider, are summarized. A few concrete selected cases are reviewed including searches for axion-like particles, anomalous $\tau$ electromagnetic moments, magnetic monopoles, and dark photons. Expectations for the achievable sensitivities of these searches in the coming years are given. Studies of CP violation in hot and dense QCD matter and connections to ultrahigh-energy cosmic rays physics are also mentioned.
\end{abstract}

\clearpage

\section{Introduction}

A wide range of experimental and observational facts --- such as \eg\ the existence of dark matter (DM), the origin of matter-antimatter asymmetry, the generation of neutrino masses, \etc\ --- and compelling theoretical arguments --- most notably the fine tuning of the Higgs boson mass, the absence of charge-parity (CP) violation in the strong interaction, the arbitrarily large range of Yukawa couplings, \etc\ --- motivate the need for new physics beyond the current Standard Model (SM) of particle physics.
Laboratory tests of the SM and direct searches for its extensions at the energy frontier capitalize on proton-proton (p-p) collisions at high center-of-mass energies aiming at producing directly or indirectly new heavy states~\cite{CidVidal:2018eel}.
On the other hand, rather than on new physics searches, efforts in high-energy heavy-ion collisions have historically focused on studying the collective behavior of partons in the quark-gluon plasma, as a means to probe the thermodynamics of nonAbelian quantum field theories in the laboratory~\cite{Citron:2018lsq}.
In the last years, however, multiple proposals have been put forward to expand both cornerstone Large Hadron Collider (LHC) programs by exploiting heavy-ion datasets as unique and complementary means to search for new phenomena~\cite{Bruce:2018yzs}.
An increasing number of studies have been appearing that exploit collisions with heavy ions at the LHC, as well as at the future Electron Ion Collider (EIC) facility~\cite{AbdulKhalek:2021gbh}, as a means to search for phenomena beyond the Standard Model (BSM).
In particular cases, the potential production and/or detection of various new proposed particles --- such as axion-like particles (ALPs)~\cite{Bauer:2017ris,Irastorza:2021tdu}, generic long-lived particles (LLPs)~\cite{Agrawal:2021dbo}, dark photons~\cite{Holdom:1985ag}, or magnetic monopoles~\cite{Affleck:1981ag,Drukier:1982}, to mention a few --- have been shown to be enhanced in collisions with ions compared to their proton-proton counterparts.

\bigskip

Most of the proposals to search for BSM physics in heavy-ion collisions are recent and form an entirely novel direction of research.
In late 2018, a first dedicated workshop on this topic \footnote{``Heavy Ions and Hidden Sectors'', \href{https://agenda.irmp.ucl.ac.be/e/Heavy-Ions-and-Hidden-Sectors}{\url{agenda.irmp.ucl.ac.be/e/Heavy-Ions-and-Hidden-Sectors}}} was organized in Louvain-la-Neuve.
The workshop brought in members of different communities (theorists and experimentalists from both the heavy-ion and BSM research areas, as well as accelerator physicists) to discuss such opportunities, and written contributions were incorporated~\cite{Bruce:2018yzs} into the discussions of the European Particle Physics Strategy Update in 2020~\cite{EuropeanStrategyforParticlePhysicsPreparatoryGroup:2019qin}.
A follow-up workshop was organized in 2021 at ECT*-Trento \footnote{``Heavy Ions and New Physics'' \href{https://indico.cern.ch/e/Heavy-Ions-and-New-Physics}{\url{indico.cern.ch/e/Heavy-Ions-and-New-Physics}}} to discuss updated proposals to exploit the potential of BSM searches
either during upcoming LHC runs or in dedicated efforts at the LHC or future colliders.
This document summarizes the presentations of this later workshop and updates the discussions submitted to the European Strategy on Particle Physics~\cite{Bruce:2018yzs}, with the goal of incorporating them into the ongoing US Community Study on the Future of Particle Physics (Snowmass 2022).
We outline a few selected physics cases, mostly connected with new light and feebly-interacting particles~\cite{Agrawal:2021dbo}, discussed during the ECT* 2021 workshop that demonstrate the benefit of using ultrarelativistic heavy-ion beams to probe novel fundamental physics phenomena in the coming decade and beyond.
Fully exploiting these exciting opportunities requires synergies among experts in the accelerator, experiment, and theory communities.
Therefore, this writeup starts first by summarizing the expected LHC performances in terms of heavy-ion integrated luminosities (Section~\ref{sec:LHC}) and is then followed by a selection of physics topics whose potential BSM impact is based on the aspects highlighted below:

\paragraph{BSM via photon-photon collisions:}

Arguably, photon-photon ($\gaga$) interactions provide the most competitive BSM search mode in collisions with heavy-ions compared to protons~\cite{Shao:2022cly}.
Thanks to their dependence on the square of the ion charge ($Z^2$), interacting electromagnetic fields in ultraperipheral heavy-ion collisions (UPCs)~\cite{Baltz:2007kq,Klein:2020fmr} have fluxes many orders of magnitude larger than those accessible in p-p collisions.
Thus, in the Pb-Pb case, the $\gaga$ cross sections can be enhanced by up to a factor of $Z^4\approx 50\cdot 10^6$ compared to their p-p or $\epem$ counterparts.
With the added benefit of no pileup collisions in the same bunch crossing, UPCs with ions can thereby probe high-energy photon interactions and search for potential modifications due to new physics in a very clean (both experimentally and theoretically) environment.
Specific target searches include:

\begin{itemize}
\item The resonant production of \textbf{axion-like particles} ($\gaga \to a \to \gaga$)~\cite{Knapen:2016moh,Bauer:2017ris,Sirunyan:2018fhl,ATLAS:2020hii} on top of the light-by-light scattering ($\gaga\to\gaga$) continuum~\cite{dEnterria:2013zqi,Aaboud:2017bwk,Sirunyan:2018fhl,Aad:2019ock}.
The lower photon detection thresholds (including trigger and offline reconstruction) available at LHCb~\cite{Goncalves:2021pdc}, as well as at the proposed next-generation ALICE-3 experiment~\cite{Adamova:2019vkf}, will provide improved limits during the upcoming LHC runs.
Section~\ref{sec:ALPs} summarizes the latest developments in this field.
For lower ALP masses, the EIC provides also interesting opportunities~\cite{Liu:2021lan}.

\item Renewed interest in the poorly constrained \textbf{anomalous electromagnetic moments of the tau lepton}~\cite{Abdallah:2003xd} has risen through the exploitation of the $\gaga\to \tau\tau$ process in UPCs~\cite{Beresford:2019gww, Dyndal:2020yen}, and details are given in Section~\ref{sec:tau}.

\item Other BSM phenomena that may be probed in $\gaga$ collisions include dark photons, searched for in photon-induced collisions ($\gamma A^\prime \to\epem$) and in the decay of neutral pions ($\gaga \to \pi^0 \to \gamma A'$)~\cite{Goncalves:2020czp}, as well as tests of Born-Infeld QED~\cite{Ellis:2017edi} or of noncommutative geometries~\cite{Hewett:2000zp, Horvat:2020ycy}.
Such studies complement the exclusive two-photon production of states with weak-scale masses such as W boson pairs ($\gaga\to\,{\rm W^+W^-}$)~\cite{Khachatryan:2016mud,ATLAS:2020qfn} and the Higgs boson ($\gaga\to\,$H$\,\to \bbbar$)~\cite{dEnterria:2009cwl,dEnterria:2019jty}.
BSM benchmarks include direct searches for doubly-charged Higgs bosons ($\gaga \to \mathrm{H^{++}H^{--}}$)~\cite{Babu:2016rcr}, and DM production via slepton ($\gaga\to \tilde{\ell}\tilde{\ell}$)~\cite{Beresford:2018pbt, Harland-Lang:2018hmi} or chargino ($\gaga\to \tilde{\chi}^+\tilde{\chi}^-$)~\cite{Godunov:2019jib} mediation.
In addition, more complicated hidden sectors, \eg\ involving two-step cascades such as $\gaga \to a \to a' a'\to \mathrm{SM}$, remain so far unexplored.

\item Nonperturbative production in the strong fields generated in UPCs via the magnetic analogue of the Schwinger effect~\cite{Gould:2017zwi,Gould:2019myj} can be used to search for \textbf{magnetic monopoles}~\cite{He:1997pj}.
Recent updates with respect to the studies reported in~\cite{Bruce:2018yzs} include: (i) a better understanding of the total and differential cross sections for monopole production by the Schwinger mechanism~\cite{Gould:2019myj,Gould:2021bre}, (ii) a demonstration that this mechanism applies to composite (as well as point-like) monopoles~\cite{Ho:2021uem}, and (iii) the first search by the MoEDAL experiment at the LHC, yielding the strongest bounds so far on the mass of possible magnetic monopoles~\cite{MoEDAL:2021vix}.
This topic is covered in Section~\ref{sec:Monopoles}.

\item Searches for \textbf{lepton flavor violation} (LFV) have been carried out usually at proton deep-inelastic facilities, such as HERA, through the conversion of a beam electron into a muon or tau via the process $ep \to \ell' X$, where $\ell' = \mu, \tau$.
An observation of such processes would be a signal of new physics, \eg\ in the form of a leptoquark (LQ) particles that couple directly to leptons and quarks.
The possibility to study LQ-based LFV scenarios using the proton beams at the EIC was considered first in Refs.~\cite{Gonderinger:2010yn,Cirigliano:2021img}, but more recently also by exploiting the large nuclear photon fluxes in ALP-mediated processes~\cite{Davoudiasl:2021mjy}.
\end{itemize}
While p-p collisions at the (HL-)LHC offer higher $\gaga$ center-of-mass energies, such studies are also possible in p-A collisions (again without pileup) that can profit in addition from proton-tagging techniques using forward proton spectrometers (AFP and PPS, respectively part of the ATLAS and CMS detectors)~\cite{ATLAS:2020bxl,Cms:2018het,CMS:2021ncv}.

\paragraph{Soft BSM physics:}

Areas in which HI collisions can offer an advantage (compared to p-p) in part of the parameter space include probes of scenarios/phenomena that require the detection of soft objects, in particular with respect to backgrounds, triggers, low-$\pT$ reconstruction, and limitations due to pileup, \etc~\cite{Drewes:2018xma,Drewes:2019vjy}.
The backgrounds in HI collisions are very different from p-p collisions, as there is practically no pileup and the risk of misidentifying the primary vertex is negligible~\cite{Sirunyan:2017xku}.
At the same time, the track multiplicity in Pb-Pb collisions is only about a factor two larger than that expected in p-p collisions at the HL-LHC with $\sim200$ pileup events, and potentially even lower than in pp collisions if lighter nuclei are used~\cite{Sirunyan:2019cgy}.
Further, the lower luminosity permits to operate the LHC main detectors with very loose kinematic triggers.
As a result, searches for new particles in HI collisions can be more sensitive than in p-p collisions when the signatures have a complicated topology, mostly come with very low $\pT$ values (usually below about 10~GeV), or their displaced vertices are in the forward direction.
This has been studied for the case of searches for heavy neutral leptons~\cite{Drewes:2018xma,Drewes:2019vjy}, and present advantages also for low mass \textbf{dark photons}~\cite{Schmidt:2021hhs}.
The latter case is discussed in Section~\ref{sec:DarkPhotons} of this report.

\paragraph{Connections to other possible BSM phenomena:}

Thermal effects in the quark-gluon plasma (QGP) can also lead to new phenomena in QCD, such as possible experimental signatures of the $\mathcal{P / CP}$ violation in strong interactions (Section~\ref{sec:CP}) via various manifestations (chiral magnetic effect~\cite{Kharzeev:2007jp,Fukushima:2008xe}, chiral magnetic waves, \etc), the production of exotic QCD states, such as strangelets~\cite{Adams:2005cu} or sexaquarks~\cite{Farrar:2020zeo} as potential DM candidates, or baryon number violation due to the strong magnetic field~\cite{Ho:2020ltr}.
They can also enhance the production cross section for magnetic monopoles~\cite{Gould:2017zwi,Bruce:2018yzs}.
Moreover, in principle thermal masses in a plasma can open up new production channels for thermal-relic DM candidates~\cite{Dvorkin:2019zdi}, though the lifetime of the QGP is too short, in general, to produce them in significant amounts, unless one can benefit from the larger chemical potential present in partonic systems produced in the lab compared to the early universe.

High-energy collisions of heavy-ions are also of relevance to understand anomalies observed in interactions of ultrahigh-energy cosmic rays (UHECR) with nuclei in the atmosphere at energies (well) above the LHC range.
Resolution of \eg\ the ``muon puzzle'' observed in such collisions~\cite{EAS-MSU:2019kmv,Pierog:2020ghc,Albrecht:2021cxw} requires to reduce the nuclear uncertainties present in the hadronic models that are used today~\cite{dEnterria:2011twh} to describe the dominant p-Air (or Fe-Air) interactions.
The possibility of an additional hard source of muons due to the early production and decay of BSM particles, such as \eg\ $(B+L)$-violating sphalerons or new heavy gauge (Z') or scalar (h) bosons, have been suggested~\cite{Brooijmans:2016lfv,Schichtel:2019hfn}.
Dedicated runs of p-O~\cite{Citron:2018lsq} and light-ion collisions at the LHC are required to improve the modelling and tuning of all nuclear effects in the current hadronic MC simulations, before one can consider any BSM interpretation of UHECR anomalies.
Section~\ref{sec:CR} discusses also the importance of understanding heavy-quark (in particular, top quark) production in ultrarelativistic HI collisions.


\section{Future performance of the LHC with heavy ions
\label{sec:LHC}}

The present official schedule of the LHC foresees a continuation of the heavy-ion program throughout Run 3, scheduled to take place in 2022--2025, and Run 4, presently planned for 2029--2032, with mainly one-month runs of Pb-Pb or p-Pb collisions per year.
A detailed baseline operational scenario has been worked out~\cite{bruce20_HL_ion_report}, relying on an improved production scheme in the injectors that allows reducing the bunch spacing to 50~ns from the previously used 75~ns~\cite{Coupard:2016vet}, as well as upgrades of the collimation system~\cite{Redaelli:2020mld,DAndrea:2021lbr}.
A range of LHC filling patterns has been worked out, with different distribution of the collisions between, on the one hand, the ALICE, ATLAS, and CMS experiments, and on the other, the LHCb experiment~\cite{bruce20_HL_ion_report}.
Therefore, some 50~ns bunch trains have to be displaced longitudinally to obtain collisions at LHCb, with the consequence of fewer collisions at the other experiments.
It is presently planned to increase the beam energy from 6.37~$Z$~TeV (corresponding to a center-of-mass energy per nucleon pair of $\sqrtsnn=5.02$~TeV, used previously in Run~2) up to 6.8~$Z$~TeV in Run~3 ($\sqrtsnn=5.36$~TeV), and likely further up to 7~$Z$~TeV in Run~4 ($\sqrtsnn=5.52$~TeV).

The expected integrated luminosity has been calculated using detailed simulations~\cite{Bruce:2021hii}, and results are shown in Table~\ref{tab:LHC_lumi}.
In a typical future one-month run, between 2.2--2.8~\nb can be expected at ALICE, ATLAS, and CMS, depending on the filling pattern, while at most five times less ($< 0.5$~\nb) would be collected at LHCb.
Five Pb-Pb runs in total until the end of Run~4 would accumulate 11--14~\nb at ALICE, ATLAS and CMS, and up to 2.5~\nb at LHCb.
For p-Pb, the projected future performance per month of operation is 470--630~\nb at ATLAS and CMS, 310--330~\nb at ALICE, and up to 170~\nb at LHCb~\cite{bruce21_evian}.
In the present planning, two such p-Pb runs are expected before the end of Run~4.
Significant uncertainties apply for both operational modes, since the nominal Pb-beam parameters in the LHC are still to be demonstrated.
The results depend also strongly on the operational efficiency, assumed at 50\%.
This factor accounts for downtime and unavailability of the LHC, faults and premature beam dumps, nonideal turnaround time, and the ramp-up period needed to reach nominal beam intensity.
Because of the short running periods, the heavy-ion runs are particularly sensitive to this latter factor, and the gains from longer runs, if any, will therefore likely be larger than proportional to allocated running time.

\begin{table}
\centering
\caption{
Projected integrated luminosity at each LHC experiment during a typical future 1-month run with Pb-Pb or p-Pb operation~\cite{Bruce:2021hii}.
Assuming five 1-month runs with Pb-Pb and two with p-Pb, the total integrated luminosity projected until the end of Run~4 in the present baseline scenario is also listed.
An operational efficiency of 50\% is assumed, as well as 24 days available for physics operation after the initial commissioning.
} \label{tab:LHC_lumi}
\tabcolsep=4.5mm
\begin{tabular}{rlccc} \toprule
 & runtime & ATLAS/CMS & ALICE & LHCb \\ \midrule
\multirow{2}{*}{Pb-Pb} & 1 month & 2.1--2.5~\nb & 2.5--2.8~\nb & $<0.5$~\nb \\
& 5 months & 10.5--12.5~\nb & 12.5--14.0~\nb & $<2.5$~\nb \\ \cmidrule{2-5}
\multirow{2}{*}{p-Pb}& 1 month & 470--630~\nb & 310--330~\nb & $<170$~\nb \\
& 2 months & 940--1260~\nb & 620--660~\nb & $<340$~\nb \\
\bottomrule \end{tabular}
\end{table}

Some improved, nonbaseline machine configurations have been studied in simulations, allowing incremental performance improvements~\cite{Bruce:2021hii}: decreasing $\beta^*$, decreasing crossing angles, and increasing the proton intensity in p-Pb runs.
Some potential minor gains could be achieved also through increased operational efficiency or if the injectors were to achieve a brightness beyond the nominal specification.
Beam studies are needed in order to conclude on the feasibility of these improvements, which could typically bring gains in integrated luminosity of up to tens of percent, or even up to a factor~2 in certain cases (\eg, for LHCb in p-Pb runs with significantly higher proton intensity).
Because of the large total cross section, the integrated luminosity is mainly limited by the total intensity that can be injected, and there is a limit on how much the turnaround time can be compressed~\cite{Bruce:2021hii}.

In addition to the baseline program using Pb-beams, a one-week pilot run with O-O and p-O collisions is planned to take place during Run~3.
The detailed operational scenario, worked out in Ref.~\cite{Bruce:2021hjk}, relies on low-intensity beams in order to simplify the validation and start production as soon as possible.
The projected integrated luminosity is about 0.5~\nb for O-O and 1.5~\nb for p-O.
The O run is motivated mainly by a physics interest (mostly to help understand the so-called ``muon puzzle'' in cosmic-ray physics~\cite{Albrecht:2021cxw}, see Section~\ref{sec:CR}), but also to some extent by the potential to gain experience with different ion species in the injector complex and LHC, in view of a possible extension of the LHC ion program beyond Run~4.
Such an extension, proposed in~\cite{Citron:2018lsq}, has the main aim of reaching a significantly higher integrated nucleon-nucleon luminosity than in Run~3 and Run~4.
As no obvious large gain factors in the Pb-Pb luminosity are within reach, collisions of other ion species are being studied.
Some first estimates of the achievable luminosity were presented in~\cite{Citron:2018lsq}, however, they did not include realistic brightness limitations from the combined effect of space charge and intrabeam scattering in the injectors and are therefore considered too optimistic.
Work is ongoing to refine the projected intensity and luminosity for a range of different ion species.


\section{Searches for axion-like particles
\label{sec:ALPs}}

Axion-like particles (ALPs) emerge as pseudo Nambu-Goldstone bosons of a new spontaneously broken global symmetry in many BSM scenarios like supersymmetry, Higgs extensions and composite dynamics models~\cite{Bauer:2017ris,dEnterria:2021ljz}.
Light pseudoscalars have been also proposed as promising dark matter candidates or dark-sector mediators.
In many scenarios, ALPs couple to photons via the effective Lagrangian:
\begin{equation}
\mathcal L = -\frac{1}{4}g_{a\gamma}\,a\,F^{\mu\nu}\tilde F_{\mu\nu}\,,
\end{equation}
where $a$ is the ALP field, $F^{\mu\nu}$ is the photon field strength tensor, and $g_{a\gamma}=1/{\rm \Lambda}_a$ is the dimensional ALP-$\gamma$ coupling constant related to the high-energy scale $\rm \Lambda$ associated with the broken symmetry.
In this scenarios, the production and decay rates of ALPs are fully defined in the parameter space of the axion mass $m_a$ and the $g_{a\gamma}$ coupling.

Ultraperipheral collisions of heavy ions provide a clean environment for ALP searches in the $m_a\lesssim 100$~GeV range~\cite{Knapen:2016moh}.
From the experimental point of view, the final state of interest for ALP production is a pair of photons emitted almost back-to-back in an otherwise empty detector.
The ALP signal would be visible as a peak in the diphoton invariant mass distribution on top of various background processes, mostly light-by-light scattering continuum, combinatorial background from $\pi^0\pi^0$ photoproduction and decay photons from spin-0 and spin-2 resonances~\cite{Klusek-Gawenda:2013rtu}.
Dedicated trigger strategies and high photon efficiency at low photon transverse momenta are crucial for a successful measurement.

\begin{figure}
\centering
\includegraphics[width=.59\linewidth]{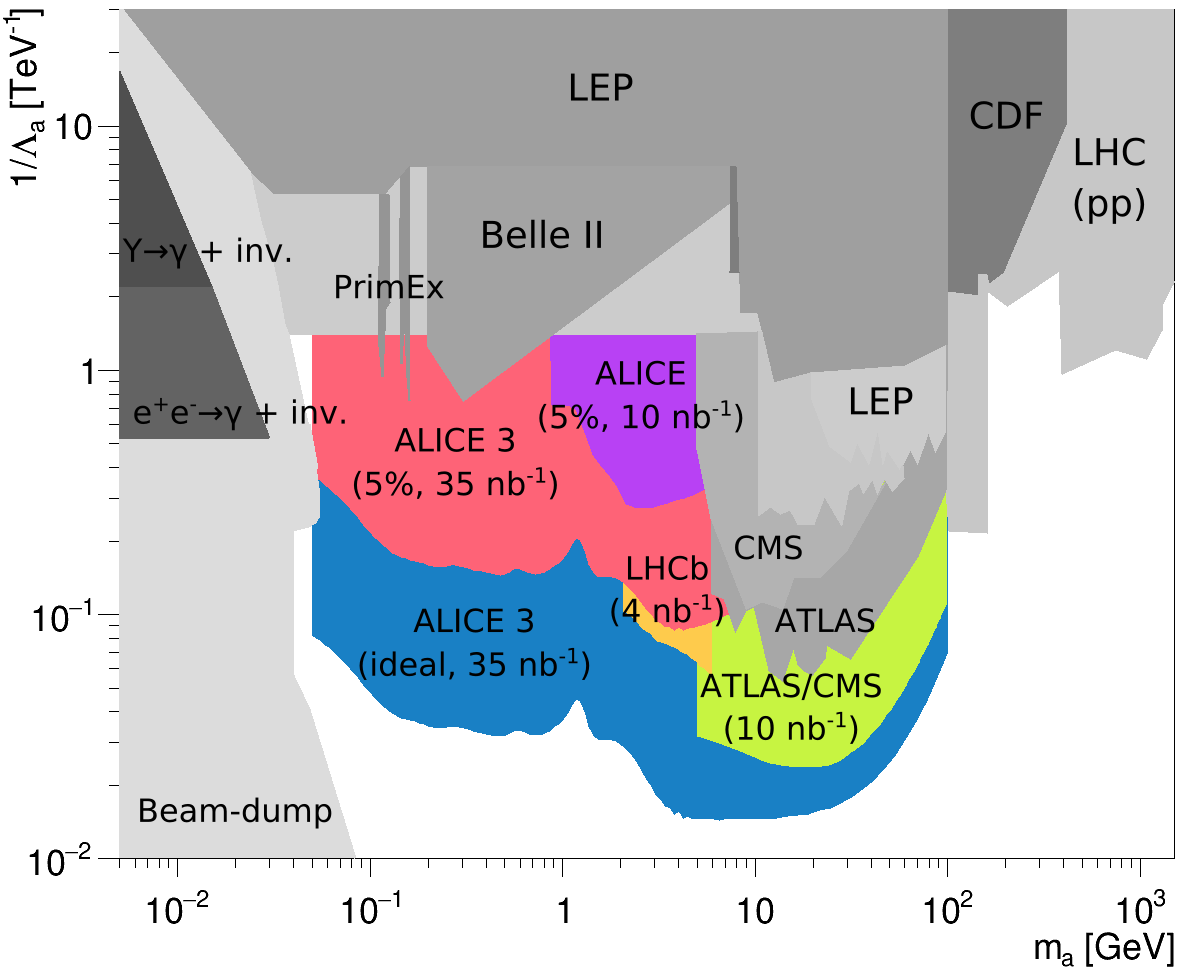}
\caption{
Bounds in the ($m_a$, $1/{\rm \Lambda}_a$) plane from existing (gray) and future (colored, with heavy-ions at the LHC) ALP searches.
} \label{fig:bsm:alps}
\end{figure}

The most stringent limits on ALPs in the $m_a$ range from $5$ to 100~GeV have been set in light-by-light scattering measurements performed in Pb-Pb UPCs at $\sqrtsnn = 5.02$~TeV by the CMS~\cite{CMS:2018erd} and ATLAS~\cite{ATLAS:2020hii} collaborations.
The ALICE and LHCb experiments can potentially improve those limits with future Run-3 and Run-4 data samples in a mass region $m_a \simeq 1\text{--}5~{\rm GeV}$ of otherwise difficult experimental access~\cite{Goncalves:2021pdc}.
Also ATLAS and CMS could extend the limits in the range $m_a<5$~${\rm GeV}$ provided a triggering strategy, and low-energy photon reconstruction are improved in future heavy-ion data taking.
The future ALICE~3 experiment~\cite{Adamova:2019vkf}, a proposed next-generation heavy-ion experiment for Run~5 and beyond, has an opportunity to extend the coverage to even lower ALP masses.
The expected performance of LHC experiments~\cite{Goncalves:2021pdc,Knapen:2016moh} is shown in Fig.~\ref{fig:bsm:alps} together with existing limits (from~\cite{ATLAS:2020hii}) and future ALICE~3 constraints estimated in two scenarios of 5\% and 100\% photon reconstruction efficiency corresponding to photons registered via conversions or in an ideal calorimeter.
As can be seen in the figure, the ALICE~3 experiment is expected to fill the gap between beam-dump and ATLAS/CMS constraints and push the limits on ALP-$\gamma$ coupling well below the $1~{\rm TeV}^{-1}$ range for intermediate masses $50~{\rm MeV}$ to $5~{\rm GeV}$, which is particularly interesting as ALPs found in this range could potentially explain the muon anomalous magnetic moment puzzle~\cite{Marciano:2016yhf,Bauer:2017ris}.

\section{Searches for anomalous electromagnetic moments of the $\tau$ lepton
\label{sec:tau}}

Ultraperipheral collisions of heavy-ions at the LHC provide a highly interesting opportunity to study the electromagnetic properties of the $\tau$ lepton via the exclusive PbPb\,$\xrightarrow{\gaga}$\,PbPb\,$\tau\tau$ process.
So far, the strongest experimental constraints on the anomalous magnetic moment of the $\tau$ lepton ($a_{\tau}$) come from the kinematics of the similar production process, $\epem \xrightarrow{\gaga} \epem\tau\tau$, measured by the DELPHI collaboration at the LEP2 collider~\cite{Abdallah:2003xd}.
Measuring $a_{\tau}$ with improved precision probes the $\tau$ lepton compositeness, and is sensitive to various BSM physics scenarios including supersymmetry~\cite{Martin:2001st}, TeV-scale leptoquarks~\cite{Feruglio:2018fxo, Crivellin:2021spu}, left-right symmetric models~\cite{GutierrezRodriguez:2004ch}, and unparticles~\cite{Moyotl:2012zz}.

In recent studies~\cite{Dyndal:2020yen,Beresford:2019gww}, it has been proposed to exploit existing and future heavy-ion datasets recorded by ATLAS and CMS experiments at the LHC for searches for $\tau$ anomalous electromagnetic moments. The $\gaga\to\tau\tau$ candidate events can be selected by requiring at least one $\tau$ lepton to decay leptonically, as this profits from the existing trigger algorithms of the ATLAS and CMS detectors.
It should be noted that the production cross section of $\tau$ lepton pairs peaks at relatively low energy/transverse momentum. Therefore, the standard $\tau$ identification tools (as developed in ATLAS/CMS for high-$\pT$ tau's) are not expected to be directly applicable.
It has been therefore proposed to categorize the $\gaga\to\tau\tau$ candidate events by their decay mode.

Possible background processes which could fake the $\gaga\to\tau\tau$ signal include the two-photon quark-antiquark production ($\gaga\to q\bar{q}$) and the exclusive production of electron/muon pairs ($\gaga\to\ell^+\ell^-$).
The $\gaga\to q\bar{q}$ processes have a significantly larger charged-particle multiplicity than the signal, and hence this background is fully reducible by exclusivity requirements.
To suppress $\gaga\to\ell^+\ell^-$ backgrounds, additional requirements on the lepton+track system transverse momentum can be applied.
A further possible sources of background are photonuclear events and semicoherent dilepton production ($\gamma^{*}\gamma\to\ell^+\ell^-$).
A requirement of zero neutrons in both ion directions, as detectable in the Zero Degree Calorimeter systems, provides a straightforward way to estimate, and even fully suppress, such backgrounds.

Systematic uncertainties are expected to be dominated by modeling of the photon flux. Here one can use a control sample of $\gaga\to\ell^+\ell^-$ events to constrain such systematics, or even eliminate them in a ratio analysis.
It is predicted that by studying the existing ATLAS/CMS Run-2 Pb-Pb dataset, one can improve the current limits on ($a_{\tau}$) by a factor of 2--3~\cite{Dyndal:2020yen,Beresford:2019gww}, thus significantly constraining many BSM physics scenarios.

A first observation of the $\gaga \to \tau\tau$ process has been recently presented at more than five $\sigma$ level by the ATLAS and CMS experiments~\cite{ATLAS:2022ryk,CMS:2022arf}, using 
Pb-Pb data collected at a center-of-mass energy per nucleon of 5.02 TeV. The cross section for this process, measured in a fiducial phase space, is consistent with leading-order QED calculations.

The ALICE experiment provides an opportunity to extend the measurement of $\gaga \to \tau\tau$ events down to low transverse momenta of leptons from $\tau$ decays in the pseudorapidity range $|\eta|<0.9$~\cite{Burmasov:2022gnl}.
During the LHC Run 3 and Run 4, ALICE plans to collect data in the continuous readout mode, and accumulate an integrated luminosity of about 2.7~nb$^{-1}$ per one month of Pb-Pb data taking.
Energy loss measurements in the ALICE Time Projection Chamber can be used for charged-particle identification and selection of leading electrons with $p_{\rm T}>0.3$~GeV from one tau decay accompanied by one or three charged tracks from the opposite-sign tau decay.
The triggerless data-taking will allow ALICE to explore a kinematic region that is difficult to access by the ATLAS and CMS experiments, and to collect an order of magnitude larger $\gaga \to \tau\tau$ data samples.
Besides, the low $p_{\rm T}$ region offers better sensitivity to negative $a_\tau$ values, thereby providing complementary information to ATLAS/CMS measurements.

\begin{figure}
\centering
\includegraphics[width=0.75\linewidth]{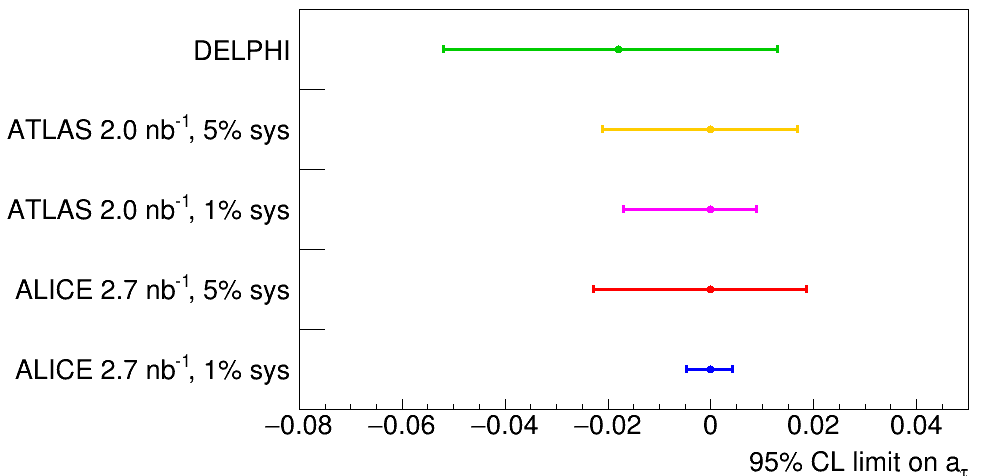}
\caption{
Expected 95\% C.L.~limits on $a_\tau$ measurements with available ATLAS/CMS Run 2 data ($2{\rm\,nb}^{-1}$) and with future ALICE data to be collected in 2022 ($2.7{\rm\,nb}^{-1}$) for two different assumptions on systematic uncertainty (1\%, 5\%) in comparison with DELPHI results~\cite{Abdallah:2003xd}.
} \label{fig:a_tau_limits}
\end{figure}

Projections on $a_\tau$ measurements with the existing ATLAS/CMS Run-2 dataset~\cite{Dyndal:2020yen} and with future ALICE data expected after one month of Pb-Pb data taking~\cite{Burmasov:2022gnl} are shown in Fig.~\ref{fig:a_tau_limits}.
The precision of these measurements can be improved by collecting more data during LHC Run-3 and Run-4, although further progress might be limited by the level of systematic uncertainties.

\section{Searches for magnetic monopoles produced via the Schwinger mechanism
\label{sec:Monopoles}}

Heavy-ion collisions offer a unique probe of the existence of magnetic monopoles (MMs), due to the strong, coherent magnetic fields produced in UPCs.
The Schwinger mechanism~\cite{Schwinger:1951nm} predicts the spontaneous creation of electron-positron pairs in the presence of a strong electric field via quantum-mechanical tunneling.
If MMs exist, the magnetic counterpart of this process~\cite{Affleck:1981ag}, via electromagnetic duality, would produce isolated magnetically charged particles in sufficiently strong magnetic fields.
The Pb-Pb collisions at the LHC generate the strongest coherent magnetic fields in the known Universe~\cite{Huang:2016}, with a peak value around $10^{20}$~G, four orders of magnitude higher than the strongest known astrophysical magnetic fields, around neutron stars.
The MM production cross section for the Schwinger mechanism can be calculated semiclassically~\cite{Gould:2017zwi,Gould:2019myj}, evading the breakdown of perturbation theory due to the strong monopole-photon coupling~\cite{Dirac:1931kp}.
Both pointlike Dirac MMs and composite MMs can be produced by the Schwinger mechanism, with the latter somewhat enhanced~\cite{Ho:2019ads,Ho:2021uem}, making it the only known mechanism capable of producing composite MMs at colliders.
Other collider searches for MMs, such as those based on p-p collision data, focus solely on pointlike Dirac MMs, due to an overwhelming exponential suppression factor for production of composite monopoles in single particle collisions~\cite{Witten:1979kh,Drukier:1982}.

The MoEDAL experiment at the LHC has conducted the first search for MMs produced in heavy-ion collisions via the Schwinger mechanism~\cite{MoEDAL:2021vix}.
The MoEDAL detectors, the Magnetic Monopole Trappers (MMTs) and the Nuclear Track Detectors, were exposed to the 2018 Pb-Pb collisions at the LHC at $\sqrtsnn = 5.02~{\rm TeV}$ and an integrated luminosity of 0.235~nb$^{-1}$.
The MMTs consist of 880~kg of Al in the form of trapping bars placed in the region around IP8.
Due to the large anomalous magnetic moment of a $^{27}_{13}$Al nucleus (100\% natural abundance), it would bind a magnetically-charged particle with an energy of 0.5--2.5 MeV~\cite{Milton:2006cp,MoEDAL:2014ttp}.
Note that there is no magnetic field in the region that would affect the MM trajectories.
After exposure, the MMT bars were scanned for trapped magnetic charges with a DC SQUID long-core magnetometer installed at the ETH Zurich Laboratory for Natural Magnetism.

\begin{figure}
\includegraphics[width=0.49\linewidth]{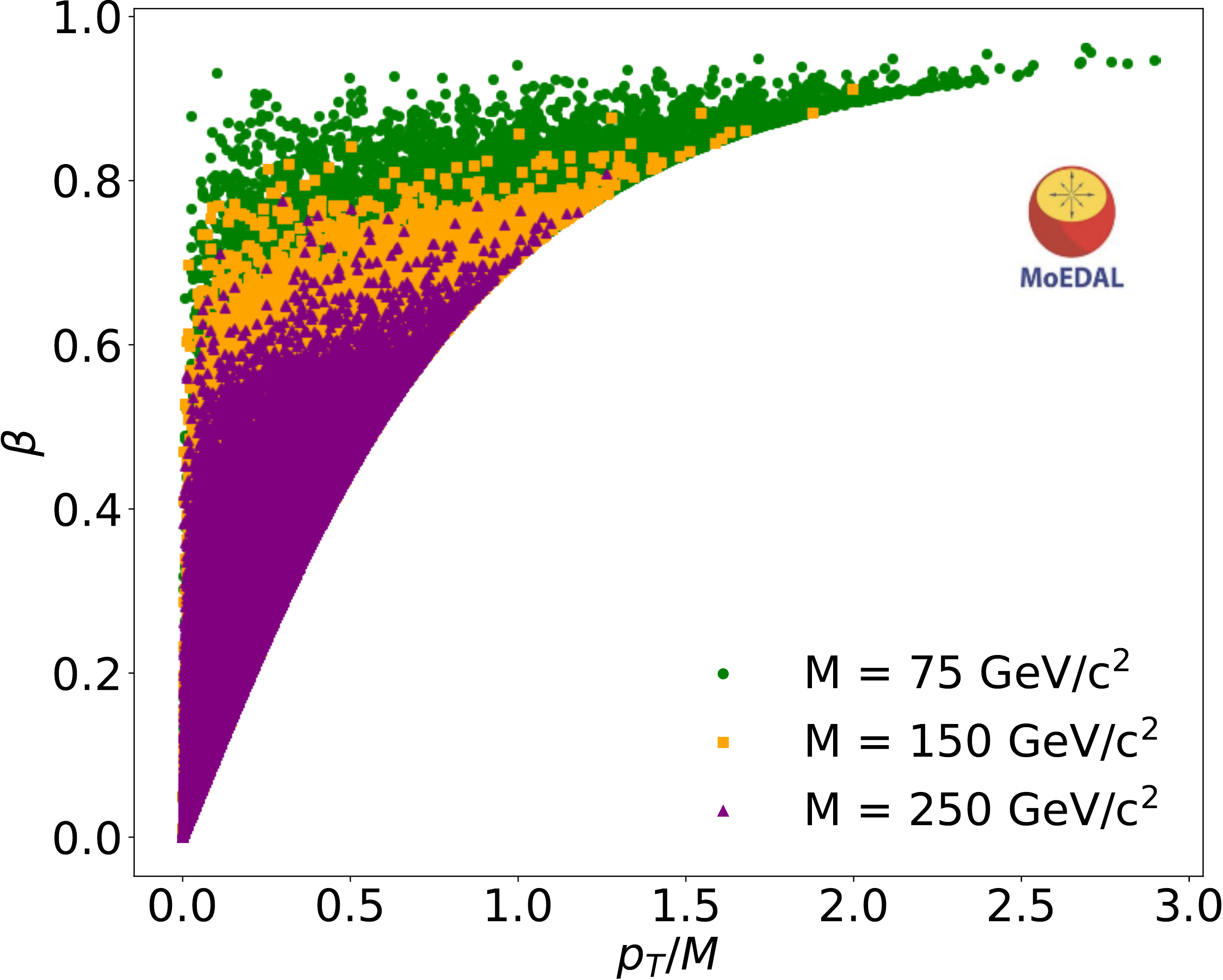}
\hfill
\includegraphics[width=0.49\linewidth]{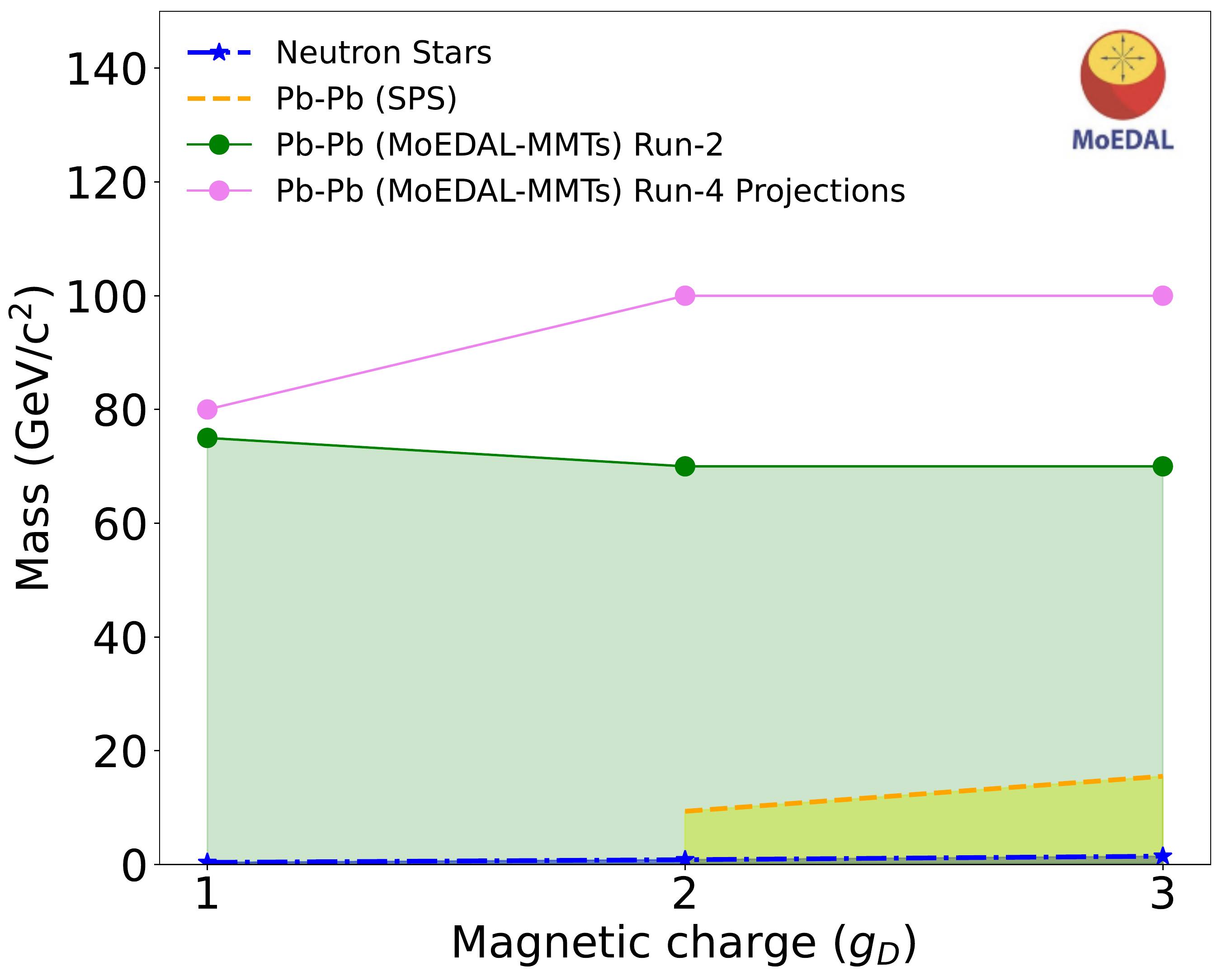}
\caption{
Left: Transverse momentum distribution for Schwinger MMs derived from the free-particle approximation, as a function of MM mass (M) plotted versus MM velocity $\beta$.
Right: The 95\% C.L.~mass exclusion regions obtained from MoEDAL's search in Pb-Pb collisions at the LHC~\cite{MoEDAL:2021vix}.
The conservative exclusion region is shaded in green.
The projected limits for future Pb-Pb runs using the conservative Schwinger cross section and the MoEDAL-MMTs are also shown.
} \label{fig:monopoles}
\end{figure}

The kinematics of MMs produced by the Schwinger mechanism may be simulated by using the free-particle approximation approach~\cite{Gould:2021bre}.
Figure~\ref{fig:monopoles} (left) shows the transverse momentum distribution for MMs, as a function of MM mass (M) plotted versus MM velocity $\beta$.
The MM ionization energy losses, geometry and material content of the MoEDAL detector is modelled~\cite{King:2016pys} in \textsc{Gauss}~\cite{Clemencic:2011zza}, which is the LHCb simulation framework that uses \textsc{Geant4} as the simulation engine.
Simulated MMs were propagated through MoEDAL detectors and the trapping efficiency, defined as the ratio of the number of MMs trapped by MMTs to the total number of generated MMs, was calculated.
The nondetection of MMs by MoEDAL~\cite{MoEDAL:2021vix}, in the first search utilising the Schwinger mechanism, resulted in the strongest bounds on the mass of possible MMs, excluding MMs with masses up to 75~GeV~\cite{MoEDAL:2021vix} as shown in Figure~\ref{fig:monopoles} (right).
The theoretical assumptions entering in the search analysis were conservative, presenting opportunities for improved theoretical analyses to extend the mass bounds in future.
In particular, the total cross section underlying Ref.~\cite{MoEDAL:2021vix} and the projections in Figure~\ref{fig:monopoles} (right) were taken to be the smallest of two different approximations, each of which is expected to provide an approximate lower bound on the true cross section~\cite{Gould:2019myj}.
The Nuclear Track Detectors in the MoEDAL-IP8 are currently being analysed for Pb-Pb Run-2 collisions with results expected in the near future.
Run-4 projections for MoEDAL's MMTs are shown in Figure~\ref{fig:monopoles} (right), highlighting the scope of future searches to extend the mass reach.

\section{Searches for dark photons
\label{sec:DarkPhotons}}

Dark photons (DPs), also called $U$-bosons or `hidden' photons $A^\prime$, are one of the possible candidate particles proposed as DM mediators.
They are supposed to interact with the SM particles via a `vector portal' due to the $U(1)-U(1)^\prime$ gauge symmetry group mixing~\cite{Holdom:1985ag}, which might make them visible in collisions of elementary particles and/or heavy ions.
The corresponding Lagrangian is defined by the hypercharge field-strength tensor of the SM photon field and the DM vector boson field: ${\cal L} \sim \epsilon^2/2 \, F_{\mu\nu}{F^{\mu\nu}}^\prime$.
Here $\epsilon^2$ is a kinetic mixing parameter, which characterizes the strength of the interaction of SM and DM particles~\cite{Fayet:1980ad,Fayet:2004bw,Boehm:2003hm,Pospelov:2007mp,Batell:2009di,Batell:2009yf}.
This mixing allows for the decay of $U$-bosons to a pair of leptons, $U\to\epem$ or $\mu^+\mu^-$.

A recent paper~\cite{Xu:2022qme} has proposed to search for $A^\prime$ in UPCs with heavy ions via photon-darkphoton collisions that produce an exclusive pair of electrons, $\gamma A^\prime\to\epem$. The existence of such a production channel would appear as an enhancement in the experimental cross section for exclusive dielectrons, as well to change in the azimuthal angular distributions, with respect to the expectations based on the Breit--Wheeler cross section for $\gaga\to\epem$~\cite{STAR:2019wlg}. From the expected data to be collected by the STAR experiment at RHIC, limits on the $m_{A'}$ vs. $\epsilon^2$ can be set, which will be the most competitive over $m_{A'}\approx 0.2$--1~GeV. Extensions of such a proposal to the case of UPCs at the LHC are in preparation.

Most commonly, light $U$-bosons are searched for in the decay of SM particles, \eg\ through Dalitz decays of pseudoscalar mesons, such as pions ($\pi^0\to \gamma U$) and $\eta$ mesons ($\eta \to \gamma U$), as well as in the Dalitz decay of baryonic resonances, such as $\Delta$'s ($\Delta \to N U$). Dark photons can be as well as produced in the decay of neutral pions produced in UPCs ($\gaga \to \pi^0 \to \gamma A'$)~\cite{Goncalves:2020czp}. Dalitz decays provide the opportunity to observe DPs in dilepton experiments from low (SIS) to ultrarelativistic (LHC) energies, which stimulated a lot of experimental as well as theoretical activities --- \cf the review~\cite{Battaglieri:2017aum}.
The HADES Collaboration at GSI, Darmstadt, performed an experimental DP search in dilepton experiments at the SIS18 accelerator with both proton and heavy-ion beams~\cite{Agakishiev:2013fwl}.
The HADES experiment presented an upper limit for the kinetic mixing parameter $\epsilon^2$ in the mass range of $M_U=0.02$--$0.55$~GeV based on the experimental measurements of $\epem$ pairs from p-p and p-Nb collisions at 3.5~GeV as well as Ar-KCl collisions at 1.76 $A$GeV.
Their result is consistent with the world data collected by various other experiments~\cite{Battaglieri:2017aum}.
Later, the HADES result has been superseded by A1~\cite{Merkel:2014avp}, NA48/2~\cite{Batley:2015lha} and BaBar~\cite{Lees:2014xha,Lees:2017lec} measurements which further lowered the limit for $\epsilon^2$ in this mass range.
The NA48/2 experiment investigated a large sample of $\pi^0$ Dalitz decays obtained from inflight weak decays of kaons, the BaBar collider experiment used their accumulated $\epem$ data sets to survey a very large mass range up to $M_U = 8$~GeV, and the MAMI-A1 experiment~\cite{Merkel:2014avp} investigated electron scattering off a $^{181}$Ta target at energies between 180 and 855~MeV to search for a DP signal.
In the mass range discussed here, $M_U = 20$--500 MeV, the limit on $\epsilon^2$ has thus been pushed down to about $10^{-6}$.
Moreover, a recent measurement of the excess electronic recoil events by the XENON1T Collaboration might be also interpreted in favor of DM sources, and in particular dark photons are possible candidates~\cite{Aprile:2020tmw}.

\begin{figure}
\includegraphics[width=.545\linewidth]{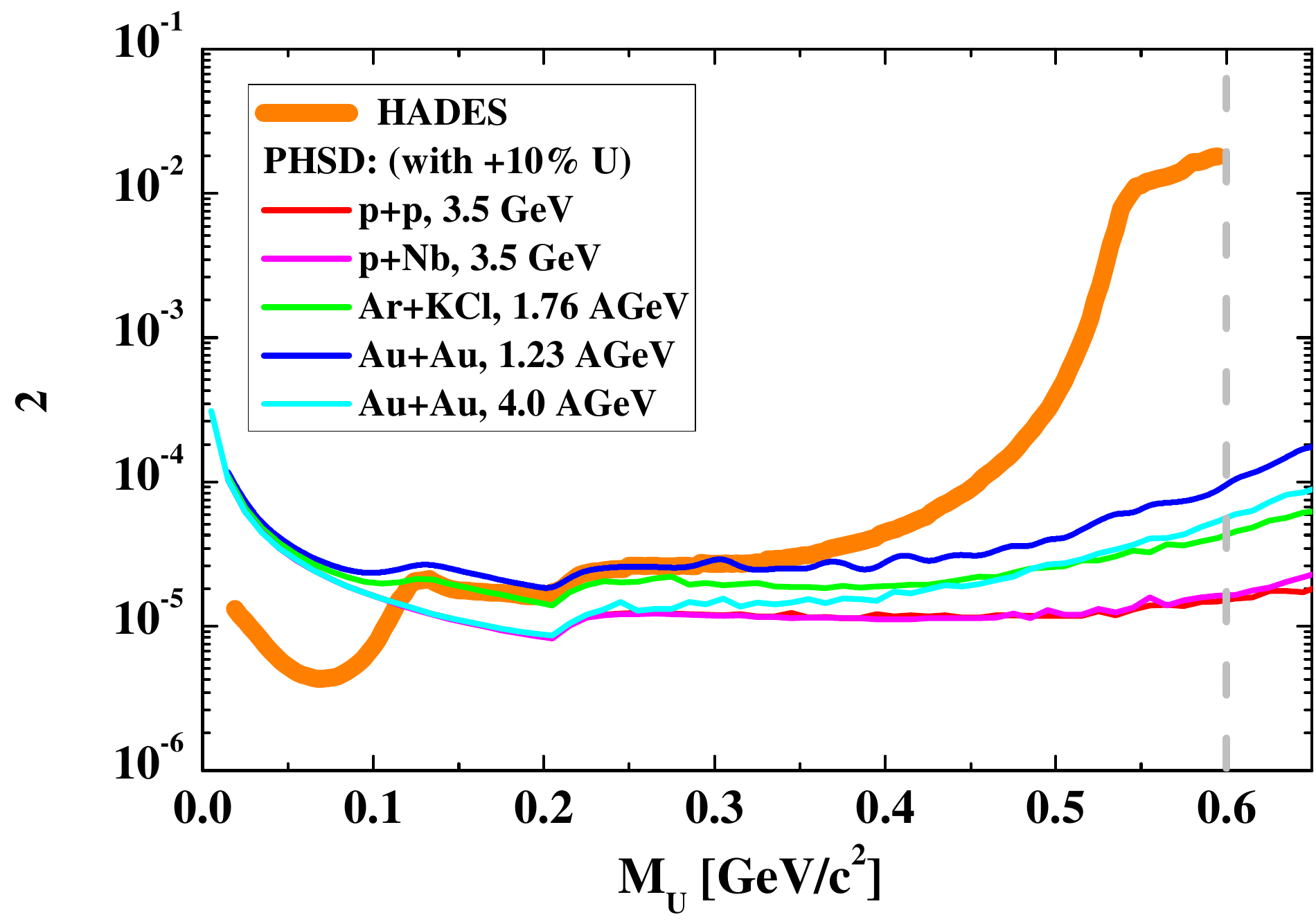}
\hfill
\includegraphics[width=.445\linewidth]{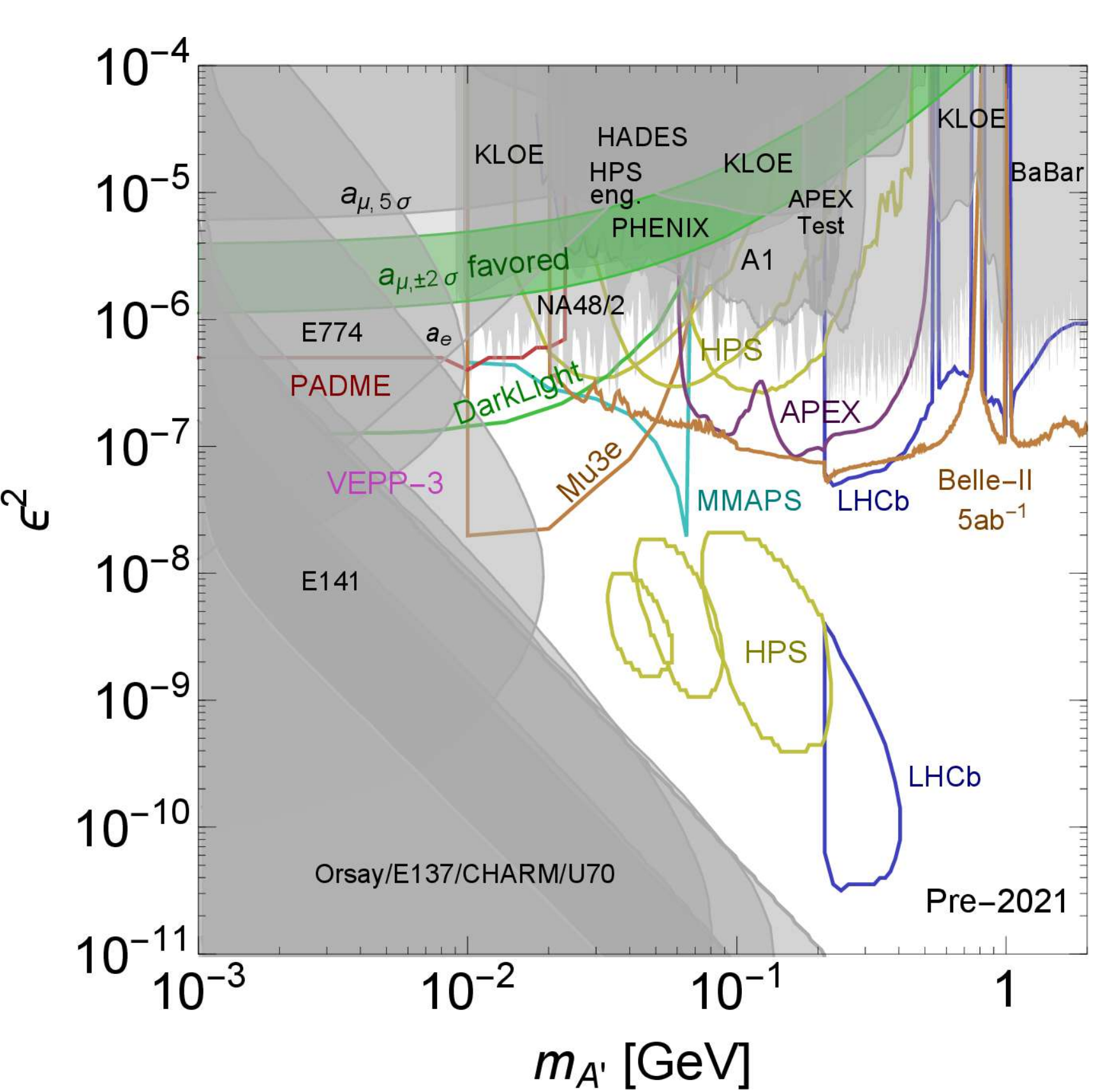}
\caption{
Left: Kinetic mixing parameter $\epsilon^2$ extracted from the PHSD dilepton spectra for p-p at 3.5~GeV (red line), p-Nb at 3.5~GeV (magenta line), Ar-KCl at 1.76~AGeV (green line), Au-Au at 1.23~AGeV (blue line), Au-Au at 4.0~AGeV in comparison with the combined HADES results (orange line)~\cite{Agakishiev:2013fwl}.
The PHSD results are shown with 10\% allowed surplus of the $U$-boson contributions over the total SM yield, \ie $C_U = 0.1$.
Right: Compilation of the experimental upper limits for $\epsilon^2$ versus the mass of dark photon set by worldwide experiments.
The figure is taken from Ref.~\cite{Battaglieri:2017aum}.
} \label{epsil2}
\end{figure}

In Ref.~\cite{Schmidt:2021hhs} a procedure to define theoretical constraints on the upper limit of $\epsilon^2(M_U)$ from the heavy-ion (as well as p-p and p-A) dilepton data has been introduced.
For that goal the light dark proton production channels have been incorporated in the microscopic Parton-Hadron-String Dynamics (PHSD) transport model~\cite{Cassing:2008sv,Bratkovskaya:2011wp,Linnyk:2015rco}, which describes the whole evolution of heavy-ion collisions based on microscopic transport theory by solving the equations-of-motion for each degree-of-freedom (partonic and hadronic) and their interactions.
The PHSD model provides a consistent description of the production of hadrons, as well as of electromagnetic probes (dileptons and photons), in p-p, p-A, and A-A collisions from SIS18 to LHC energies~\cite{Linnyk:2015rco,Song:2018xca}, \ie\ it provides a rather good estimate of the SM backgrounds in searches for dark photon radiation to dileptons.
Using the fact that the dark photons are not observed in dilepton experiments so far one can require that their contribution cannot exceed some limit which would make them visible in experimental data.
By varying the parameter $\epsilon^2(M_U)$ in the model calculations, one can obtain upper constraints on $\epsilon^2(M_U)$ based on pure theoretical results for dilepton spectra under the constraint that the `surplus' of DM contribution does not overshine the SM contributions (which is equivalent to the measured dilepton spectra) within any requested accuracy.

In Fig.~\ref{epsil2} (left) we show the results for the kinetic mixing parameter $\epsilon^2$ versus $M_U$ extracted from the PHSD dilepton spectra for p-p at 3.5~GeV (red line), p-Nb at 3.5~GeV (magenta line), Ar-KCl at 1.76~AGeV (green line), Au-Au at 1.23~AGeV (blue line), Au-Au at 4.0~AGeV in comparison with the combined HADES results (orange line) from Ref.~\cite{Agakishiev:2013fwl}.
The PHSD results are shown with 10\% allowed surplus of the $U$-boson contributions over the total SM yield.
The right plot of Fig.~\ref{epsil2} shows the compilation of the experimental upper limits for $\epsilon^2$ versus the mass of dark photon set by worldwide experiments~\cite{Battaglieri:2017aum}.
We note, that our analysis can help to estimate the requested accuracy for future experimental searches of light DPs in dilepton experiments.
This procedure can be extended for searches for DP of any masses with the corresponding production and decay channels implemented in the PHSD code.
Moreover, it can be extended for an estimate of the contribution of other dark matter candidates to the hadronic observables in heavy-ion experiments.

\bigskip

\begin{figure}
\centering
\includegraphics[width=.645\linewidth]{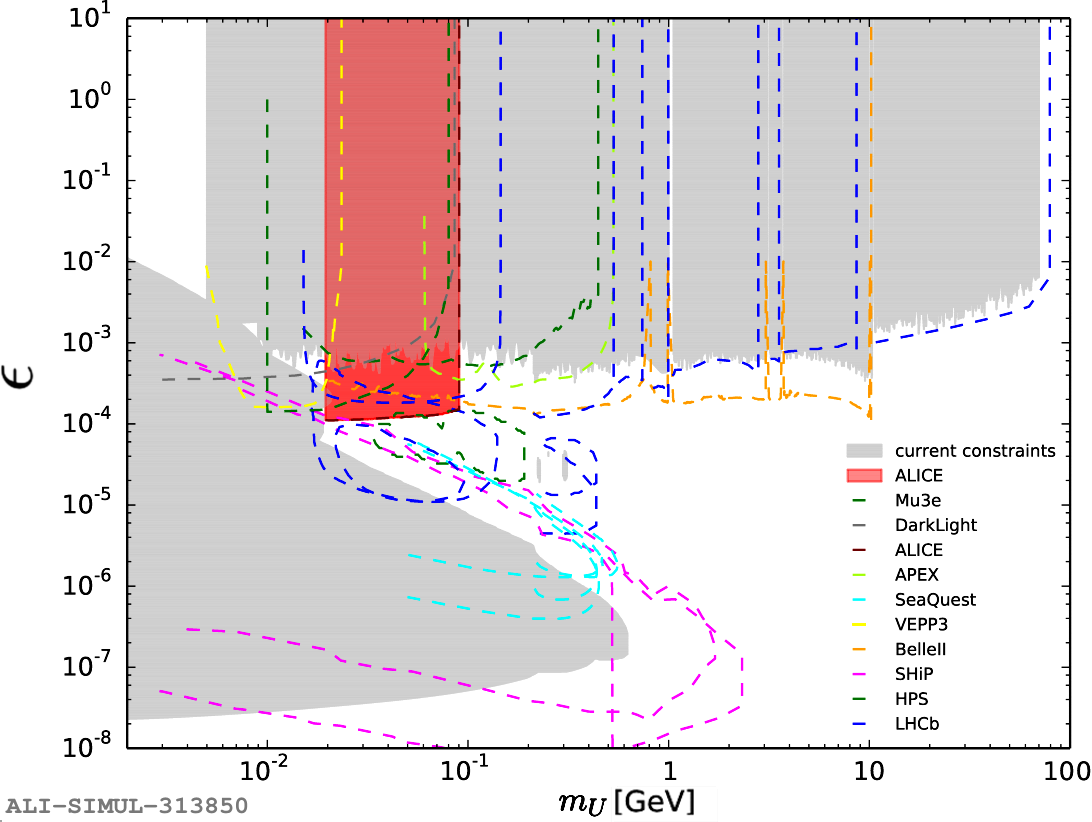}
\caption{
Dark photon limits in the coupling \vs mass plane.
The red area indicates the expected 90\% C.L.~constraints from ALICE measurements before 2030.
Figure adapted from Refs.~\cite{Citron:2018lsq,Ilten:2018crw}.
} \label{fig:DP_ALICE}
\end{figure}

ALICE, as the LHC dedicated experiment to heavy-ion physics, has good capabilities for electron identification in the low transverse momentum region, thereby enabling the measurement of a large sample of $\pi^0$ and $\eta$ Dalitz decays of relevance for DP searches~\cite{Acharya:2018ohw}.
ALICE preliminary result from Run-1 p-p data sets limits of $\epsilon^2 = 3\cdot10^{-5}$ in the mass range of 20--100 MeV.
The upgrade during LS2 will strongly improve the efficiency of e$^\pm$ measurements and data taking capabilities.
ALICE estimates that the range of couplings $\epsilon^2 \approx (2\text{--}5) \cdot 10^{-8}$ will be reached in the mass range of $M_{U} = 0.02$--0.1~GeV using data samples to be collected in Runs 3 and 4:~~6~pb$^{-1}$ of p-p, 0.3~pb$^{-1}$ of p-Pb at 0.5 T and 0.2 T magnetic field, 10~nb$^{-1}$, and 3~nb$^{-1}$ in Pb-Pb collisions at 0.5 T and 0.2 T magnetic field, respectively~\cite{Citron:2018lsq} (Fig.~\ref{fig:DP_ALICE}).
Beyond this, the future ALICE~3 experiment~\cite{Adamova:2019vkf} for Run-5 and Run-6 has a strong potential to lower the limit of $\epsilon^2(M_U)$.
Thanks to precision tracking performance down to very low $\pT$, improved tracking and mass resolution, and high efficiencies for dielectron reconstruction and identification, approximately $\times 60$ larger $\pi^0$ and $\eta$ Dalitz samples will be measured, and mixing parameters as low as $\epsilon^2 \approx (4\text{--}10) \cdot 10^{-9}$ and $\epsilon^2 \approx (1$--$6) \cdot 10^{-8}$ will be reached in the $\pi^0$ and $\eta$ mass region, respectively.

\section{Chiral effects in hot and dense QCD matter
\label{sec:CP}}

Studies of structures with nontrivial topology in the QCD vacuum, which determine the behavior of the $\mathcal{P} / \mathcal{CP}$ fundamental symmetries in hot quark-gluon matter, can shed light in our understanding of $\mathcal{P} / \mathcal{CP}$ symmetries in the strong interaction. As discussed in~\cite{Okorokov:2017pxd}, collisions of relativistic nuclei generate very strong electric $\bf{E}$ and magnetic $\bf{B}$ fields.
Owing to a nontrivial topology of the QCD vacuum, the interactions of the external extremely strong Abelian fields with the final-state QGP give rise to dissipationless transport phenomena~\cite{Landsteiner:2016led} which may lead to interesting effects observable experimentally, such as \eg\ (i) the chiral magnetic effect (CME), (ii) the chiral separation effect (CSE), (iii) the electric separation effect (ESE), and/or (iv) the emergence of the chiral vortical effect (CVE).
Superpositions of the aforementioned effects may lead to collective modes in quark-gluon matter.
A combination of CSE and the CME renders possible the existence of a new type collective mode in the QGP, such as the chiral magnetic wave (CMW).
By analogy with CMW, it was theoretically predicted chiral vortical wave (CVW) involving the collective propagation of fluctuations of the electrical- and chiral-charge densities along the vorticity $\bf{\omega}$ direction~\cite{Kharzeev:2015znc}.

A hard-sphere model of heavy-ion collisions without medium effects (electric conductivity \etc) predicts a peak value of the magnetic-field strength $eB$ about tens of GeV$^{2}$~\cite{Okorokov:2019bsi} even for collisions of light nuclei at energy ranges of the HE-LHC~\cite{FCC:2018bvk} and FCC~\cite{FCC:2018vvp,Dainese:2016gch} projects.
The absolute value of the magnetic field rapidly decreases with time and increases with atomic number, thereby providing new ways to search for experimental signatures of the nontrivial topology of the QCD vacuum.
Due to high luminosity and high multiplicity per event in the HE-LHC and FCC energy domain, various multiparticle (azimuthal) correlations techniques can be used to investigate the wide set of aforementioned chiral effects.
Also, event shape engineering (ESE) technique is considered as another
promising tool to disentangle background contributions from the possible physical signals.
The corresponding measurements require larger data samples. Furthermore, the future heavy-ion collider projects under consideration can provide the opportunity for the study of the flavor dependence of the $\mathcal{P} / \mathcal{CP}$ violation via the measured correlations of different particle species.
One can expect a mass ordering of the strength of chiral effects, in particular, of the CME.
Heavy quarks appear to be less affected by the influence of the external $\bf{B}$ than light
quarks even for the extremely strong $eB \sim 10$~GeV$^{2}$.
Therefore one can qualitatively expect a smaller magnitude of the CME for charm and bottom particles.
The measurements with heavy quarks can provide new unique information with regard to the sphaleron and the strong $\mathcal{CP}$ problem.

Preliminary measurements of 3-particle correlators in p-p, p-$\Lambda$ and p-$\pi$ final-state pairs are consistent with the presence of the CVE, whose contribution is larger than that of the CME~\cite{Zhao:2014aja}.
Investigations of the multipoint azimuthal correlations in particle ensembles with baryons can shed new light on the collective chiral modes in QGP.
Furthermore, the study of small partonic systems (produced in p-A and light nuclei collisions) at ultrahigh energies may provide clearer signals for the formation of novel vortical structures which can be generated, in general, in asymmetric collisions due to the special geometry and unique dynamics~\cite{Lisa:2021zkj}.
High luminosity measurements in collisions of up to tens TeV will allow the precise measurements of global polarization of various particle species including multistrange baryons and heavy flavors ($c$, $b$) in the presence of the QGP phase.
Such measurements would be useful for the experimental study of anomalous gravitomagnetic moment (AGM) deeply related with the axial charge separation along the rotation of the fluid~\cite{Kharzeev:2015znc} and consequent verification of the hypothesis that thermal effects can break the Einstein equivalence principle in quantum field theory~\cite{Buzzegoli:2021jeh}.

\section{Connections between ultrahigh-energy cosmic rays and heavy-ions physics
\label{sec:CR}}

Measurements of collisions of cosmic ray (CR) particles with incoming energies in the range $10^{18}$--$10^{21}$~eV with nuclei in the upper atmosphere provide new unique opportunities for the
study of multiparticle production processes at energies (well) above 
those reachable at the LHC and future colliders~\cite{dEnterria:2011twh,Kampert:2012mx}.
The energy range for protons in the lab reference frame associated with such ultrahigh energy cosmic rays (UHECR) includes the so-called Greisen-Zatsepin-Kuzmin (GZK) limit of $E_{p}\approx10^{20}$~eV~\cite{Greisen:1966jv,Zatsepin:1966jv} and somewhat expands it, taking into account possible uncertainties in the theoretical estimations of the limit energy values for UHECR and the experimental measurements~\cite{RP-333-309-2000,PU-56-304-2013}.
As shown in~\cite{Okorokov-PAN-82-838-2019}, the partonic medium produced in the central collisions of UHECR particles with air nuclei is characterized by high energy densities at midrapidity and temperatures well above the critical ones for the creation of the QGP already at $E_{p} \gtrsim 10^{17}$~eV.
The Bose-Einstein decoupling time is of the order of $\sim 10$~fm even in collisions induced by ${}^{4}\mbox{He}^{2+}$ CRs.
The particle source created in proton-nucleus collisions at $E_{p} \gtrsim 10^{19}$~eV is characterized by large space-time extents, which support the hypothesis of the creation of blobs of QGP in UHECR interactions~\cite{Okorokov-PAN-82-838-2019},
where phenomena due to partonic collectivity can take place~\cite{Kalaydzhyan:2014uha}.

One of the most remarkable features of recent UHECR measurements is the observed muon excess with respect to Monte Carlo model predictions at energies of primary particles above $\sim10^{17}$~eV~\cite{EAS-MSU:2019kmv}.
The role of QGP formation in the small partonic systems produced in UHECR energy range can be relevant for the resolving such a ``muon puzzle''~\cite{Albrecht:2021cxw}.
In particular, collective hadronization in the QGP can play a significant role for the air-shower development~\cite{Pierog:2020ghc}.
The study of the matter created in interactions of UHECR particles
with air nuclei is relevant for both the present UHECR observatories and for heavy-ion collisions (including proton-nucleus and light nuclei) at future colliders.
The upper $E_{p} \gtrsim 10^{20}$~eV boundary of the UHECR energy domain corresponds to the hundreds TeV and even $\mathcal{O}$(1 PeV) in $\sqrts$.
At such energies, the production of heavy quarks, including charm and beauty~\cite{dEnterria:2016yhy} and top~\cite{dEnterria:2015mgr} quarks, is large and opens a unique possibility for investigating the preequilibrium stages of the space-time evolution of the QGP.
Cross sections for charm and beauty production in p-Air collisions at GZK-cutoff energies reach values of $\sigma(c\bar{c}) \approx 5$~b, and $\sigma(b\bar{b}) \approx 100$~mb, respectively.
For top quarks, the first evidence for their production in heavy-ion interactions has been recently presented by CMS in Pb-Pb collisions at $\sqrtsnn=5.02$~TeV~\cite{arxiv-2006.11110-2020}.
The sum of pQCD NNLO partonic cross sections for $t\bar{t}$ production results in an estimated value of $\sigma(t\bar{t}) \approx 70~\text{nb}\times A^2$ at $\sqrtsnn \approx 0.5$~PeV~\cite{Okorokov-JPCS-1690-012006-2020}, implying $\sigma(t\bar{t}) \approx 1$~$\mu$b for p-Air collisions.
A recent Monte Carlo study~\cite{dEnterria:2018kcz} showed that heavy-quark ($c$, $b$) production, as implemented for p-p collisions, cannot explain alone the muon puzzle in extended atmospheric showers, thereby confirming the need for additional nuclear effects~\cite{Pierog:2020ghc}.
However, the future detailed investigation of the production of heavy flavors, in particular top, in high-energy heavy- and light-ion collisions can provide novel insights into both fields --- collider and UHECR physics --- with interdisciplinary and cross-fertilizing aspects~\cite{Okorokov-JPCS-1690-012006-2020}.

\section{Summary}

This short report presents several recent proposals that exploit heavy-ion (HI) collider data to search for physics beyond the Standard Model (BSM), and updates and expands those of a previous paper with similar scope~\cite{Bruce:2018yzs}.
A noncomprehensive but representative list of BSM processes accessible with HI at the LHC has been presented.
Despite the lower nucleus-nucleus c.m.~energies and beam luminosities compared to p-p collisions, HI are more competitive than the latter in particular in BSM scenarios, whereas in some others they can complement or confirm searches (or discoveries) performed in the p-p mode.
The advantages of HI with respect to p-p searches are either based on comparatively enhanced underlying mechanisms of production: $\gamma\gamma$ processes in ultraperipheral collisions and ``Schwinger'' production through strong classical EM fields, or on improved reconstruction of new physics signals in the ``soft'' regime.

Topics reviewed in this paper include novel ways to search for axion-like particles and dark photons; proposals to constrain the anomalous electromagnetic moments of the tau lepton; deeper theoretical studies of magnetic monopole production in heavy ion collisions, as well as the first experimental results from the dedicated MoEDAL detector at the LHC, yielding the strongest bounds so far.
Proposals to improve our understanding of charge and parity violation in the strong interaction, and the synergy between HI collisions and the study of new phenomena in ultrahigh-energy cosmic rays have been also outlined.
The topics covered here provide additional motivations, beyond the traditional QGP/QCD physics cases, to prolong the HI program at the LHC past their currently scheduled end in 2032 (Run-4), in particular with lighter ion systems, an LHC running mode that has not been operated so far.

\subsection*{Acknowledgements}

We gratefully acknowledge the support of the ECT* Trento center for the organization of the workshop ``Heavy Ions and New Physics'' that took place virtually on 20 and 21 May 2021. In particular, support by the EU Horizon 2020 research and innovation programme, STRONG-2020 project, under grant agreement No 824093 is acknowledged.
We wish to thank all participants of the workshop, for the lively and fruitful discussions that kicked-off the writing of this document. 
The work of IGB was partly supported by the National Science Centre of Poland under grant number UMO-2020/37/B/ST2/01043.
The work of MD is cofinanced by the Polish National Agency for Academic Exchange within Polish Returns Programme, Grant No.\ PPN/PPO/2020/1/00002/U/00001.
The work of EK and NB was supported by RFBR, grant no.~21-52-14006.
The work of NB was supported by RSCF, grant no.~22-42-04405.
The work of VAO was supported partly by NRNU MEPhI Program ``Priority 2030''.
The work of AU was supported by the NSF grant 2011214

\clearpage
\printbibliography

@article{Bruce:2018yzs,
    author = "Bruce, Roderik and others",
    title = "{New physics searches with heavy-ion collisions at the CERN Large Hadron Collider}",
    eprint = "1812.07688",
    archivePrefix = "arXiv",
    primaryClass = "hep-ph",
    doi = "10.1088/1361-6471/ab7ff7",
    journal = "J. Phys. G",
    volume = "47",
    pages = "060501",
    year = "2020"
}

@TechReport{bruce20_HL_ion_report,
  author  = {R. Bruce and T. Argyropoulos and H. Bartosik and R. De Maria and N. Fuster-Martinez and M.A. Jebramcik and {J.M.~Jowett} and N. Mounet and S. Redaelli and G. Rumolo and M. Schaumann and H. Timko},
  title   = {{HL-LHC} operational scenario for {Pb-Pb} and {p-Pb} operation},
  number = {CERN-ACC-2020-0011},
  year    = {2020},
  url     = {https://cds.cern.ch/record/2722753},
}

@article{Coupard:2016vet,
    author = "Coupard, J. and Damerau, H. and Funken, A. and Garoby, R. and Gilardoni, S. and Goddard, B. and Hanke, K. and Manglunki, D. and Meddahi, M. and Rumolo, G. and Scrivens, R. and Chapochnikova, E.",
    title = "{LHC Injectors Upgrade, Technical Design Report}: {Vol. II: Ions}",
    reportNumber = "CERN-ACC-2016-0041",
    month = "4",
    year = "2016",
    url = "https://cds.cern.ch/record/2153863"
}

@article{Redaelli:2020mld,
    author = "Redaelli, Stefano and Bruce, Roderik and Lechner, Anton and Mereghetti, Alessio",
    editor = {B\'ejar Alonso, I. and Br\"uning, O. and Fessia, P. and Rossi, L. and Tavian, L. and Zerlauth, M.},
    title = "{Chapter 5: Collimation system}",
    doi = "10.23731/CYRM-2020-0010.87",
    journal = "CERN Yellow Rep. Monogr.",
    volume = "10",
    pages = "87",
    year = "2020"
}

@inproceedings{DAndrea:2021lbr,
    author = "D'Andrea, Marco and others",
    title = "{Crystal Collimation of 20 MJ Heavy-Ion Beams at the HL-LHC}",
    booktitle = "{12th International Particle Accelerator Conference~}",
    doi = "10.18429/JACoW-IPAC2021-WEPAB023",
    month = "8",
    year = "2021"
}

@article{Bruce:2021hii,
    author = "Bruce, R. and Jebramcik, M. A. and Jowett, J. M. and Mertens, T. and Schaumann, M.",
    title = "{Performance and luminosity models for heavy-ion operation at the CERN Large Hadron Collider}",
    eprint = "2107.09560",
    archivePrefix = "arXiv",
    primaryClass = "physics.acc-ph",
    doi = "10.1140/epjp/s13360-021-01685-5",
    journal = "Eur. Phys. J. Plus",
    volume = "136",
    pages = "745",
    year = "2021"
}

@unpublished{bruce21_evian,
  author  = {{R. Bruce \textit{et al.}}},
  title   = {{Plans for LHC ion operation in Run 3}},
  note = {Presentation at the 10th LHC Operations Workshop, November 2021, CERN, Geneva, Switzerland},
  year    = {2021},
  url     = {https://indico.cern.ch/event/1077835/contributions/4533358},
}

@inproceedings{Battaglieri:2017aum,
    author = "Battaglieri, Marco and others",
    title = "{US Cosmic Visions: New Ideas in Dark Matter 2017: Community Report}",
    booktitle = "{U.S. Cosmic Visions: New Ideas in Dark Matter}",
    eprint = "1707.04591",
    archivePrefix = "arXiv",
    primaryClass = "hep-ph",
    reportNumber = "FERMILAB-CONF-17-282-AE-PPD-T",
    month = "7",
    year = "2017"
}

@article{Citron:2018lsq,
    author = "Citron, Z. and others",
    editor = "Dainese, Andrea and Mangano, Michelangelo and Meyer, Andreas B. and Nisati, Aleandro and Salam, Gavin and Vesterinen, Mika Anton",
    title = "{Report from Working Group 5}: {Future physics opportunities for high-density QCD at the LHC with heavy-ion and proton beams}",
    eprint = "1812.06772",
    archivePrefix = "arXiv",
    primaryClass = "hep-ph",
    reportNumber = "CERN-LPCC-2018-07",
    doi = "10.23731/CYRM-2019-007.1159",
    journal = "CERN Yellow Rep. Monogr.",
    volume = "7",
    pages = "1159",
    year = "2019"
}

@inproceedings{Bruce:2021hjk,
    author = "Bruce, Roderik and Alemany-Fern\'andez, Reyes and Bartosik, Hannes and Jebramcik, Marc and Jowett, John and Schaumann, Michaela",
    title = "{Studies for an LHC Pilot Run with Oxygen Beams}",
    booktitle = "{12th International Particle Accelerator Conference~}",
    doi = "10.18429/JACoW-IPAC2021-MOPAB005",
    month = "8",
    year = "2021"
}

@article{Holdom:1985ag,
    author = "Holdom, Bob",
    title = "{Two U(1)'s and Epsilon Charge Shifts}",
    reportNumber = "UTPT-85-30",
    doi = "10.1016/0370-2693(86)91377-8",
    journal = "Phys. Lett. B",
    volume = "166",
    pages = "196",
    year = "1986"
}

@article{Fayet:1980ad,
    author = "Fayet, Pierre",
    title = "{Effects of the Spin 1 Partner of the Goldstino (Gravitino) on Neutral Current Phenomenology}",
    reportNumber = "CERN-TH-2895",
    doi = "10.1016/0370-2693(80)90488-8",
    journal = "Phys. Lett. B",
    volume = "95",
    pages = "285",
    year = "1980"
}

@article{Fayet:2004bw,
    author = "Fayet, Pierre",
    title = "{Light spin 1/2 or spin 0 dark matter particles}",
    eprint = "hep-ph/0403226",
    archivePrefix = "arXiv",
    reportNumber = "LPTENS-04-15",
    doi = "10.1103/PhysRevD.70.023514",
    journal = "Phys. Rev. D",
    volume = "70",
    pages = "023514",
    year = "2004"
}

@article{Boehm:2003hm,
    author = "Boehm, C. and Fayet, Pierre",
    title = "{Scalar dark matter candidates}",
    eprint = "hep-ph/0305261",
    archivePrefix = "arXiv",
    doi = "10.1016/j.nuclphysb.2004.01.015",
    journal = "Nucl. Phys. B",
    volume = "683",
    pages = "219",
    year = "2004"
}

@article{Pospelov:2007mp,
    author = "Pospelov, Maxim and Ritz, Adam and Voloshin, Mikhail B.",
    title = "{Secluded WIMP Dark Matter}",
    eprint = "0711.4866",
    archivePrefix = "arXiv",
    primaryClass = "hep-ph",
    doi = "10.1016/j.physletb.2008.02.052",
    journal = "Phys. Lett. B",
    volume = "662",
    pages = "53",
    year = "2008"
}

@article{Batell:2009di,
    author = "Batell, Brian and Pospelov, Maxim and Ritz, Adam",
    title = "{Exploring Portals to a Hidden Sector Through Fixed Targets}",
    eprint = "0906.5614",
    archivePrefix = "arXiv",
    primaryClass = "hep-ph",
    doi = "10.1103/PhysRevD.80.095024",
    journal = "Phys. Rev. D",
    volume = "80",
    pages = "095024",
    year = "2009"
}

@article{Batell:2009yf,
    author = "Batell, Brian and Pospelov, Maxim and Ritz, Adam",
    title = "{Probing a Secluded U(1) at B-factories}",
    eprint = "0903.0363",
    archivePrefix = "arXiv",
    primaryClass = "hep-ph",
    doi = "10.1103/PhysRevD.79.115008",
    journal = "Phys. Rev. D",
    volume = "79",
    pages = "115008",
    year = "2009"
}

@article{Aprile:2020tmw,
    author = "Aprile, E. and others",
    collaboration = "XENON",
    title = "{Excess electronic recoil events in XENON1T}",
    eprint = "2006.09721",
    archivePrefix = "arXiv",
    primaryClass = "hep-ex",
    doi = "10.1103/PhysRevD.102.072004",
    journal = "Phys. Rev. D",
    volume = "102",
    pages = "072004",
    year = "2020"
}

@article{Agakishiev:2013fwl,
    author = "Agakishiev, G. and others",
    collaboration = "HADES",
    title = "{Searching a Dark Photon with HADES}",
    eprint = "1311.0216",
    archivePrefix = "arXiv",
    primaryClass = "hep-ex",
    doi = "10.1016/j.physletb.2014.02.035",
    journal = "Phys. Lett. B",
    volume = "731",
    pages = "265",
    year = "2014"
}

@article{Merkel:2014avp,
    author = "Merkel, H. and others",
    title = "{Search at the Mainz Microtron for Light Massive Gauge Bosons Relevant for the Muon $g-2$ Anomaly}",
    eprint = "1404.5502",
    archivePrefix = "arXiv",
    primaryClass = "hep-ex",
    doi = "10.1103/PhysRevLett.112.221802",
    journal = "Phys. Rev. Lett.",
    volume = "112",
    pages = "221802",
    year = "2014"
}

@article{Batley:2015lha,
    author = "Batley, J. R. and others",
    collaboration = "NA48/2",
    title = "{Search for the dark photon in $\pi^0$ decays}",
    eprint = "1504.00607",
    archivePrefix = "arXiv",
    primaryClass = "hep-ex",
    reportNumber = "CERN-PH-EP-2015-093",
    doi = "10.1016/j.physletb.2015.04.068",
    journal = "Phys. Lett. B",
    volume = "746",
    pages = "178",
    year = "2015"
}

@article{Lees:2014xha,
    author = "Lees, J. P. and others",
    collaboration = "BaBar",
    title = "{Search for a Dark Photon in $e^+e^-$ Collisions at BaBar}",
    eprint = "1406.2980",
    archivePrefix = "arXiv",
    primaryClass = "hep-ex",
    reportNumber = "BABAR-PUB-14-002, SLAC-PUB-15979",
    doi = "10.1103/PhysRevLett.113.201801",
    journal = "Phys. Rev. Lett.",
    volume = "113",
    pages = "201801",
    year = "2014"
}

@article{Lees:2017lec,
    author = "Lees, J. P. and others",
    collaboration = "BaBar",
    title = "{Search for Invisible Decays of a Dark Photon Produced in ${e}^{+}{e}^{-}$ Collisions at BaBar}",
    eprint = "1702.03327",
    archivePrefix = "arXiv",
    primaryClass = "hep-ex",
    reportNumber = "BABAR-PUB-17-001, SLAC-PUB-16923",
    doi = "10.1103/PhysRevLett.119.131804",
    journal = "Phys. Rev. Lett.",
    volume = "119",
    pages = "131804",
    year = "2017"
}

@article{Schmidt:2021hhs,
    author = "Schmidt, Ida and Bratkovskaya, Elena and Gumberidze, Malgorzata and Holzmann, Romain",
    title = "{Constraints on the kinetic mixing parameter \ensuremath{\varepsilon}2 for the light dark photons from dilepton production in heavy-ion collisions in the few-GeV energy range}",
    eprint = "2105.00569",
    archivePrefix = "arXiv",
    primaryClass = "hep-ph",
    doi = "10.1103/PhysRevD.104.015008",
    journal = "Phys. Rev. D",
    volume = "104",
    pages = "015008",
    year = "2021"
}

@article{Cassing:2008sv,
    author = "Cassing, W. and Bratkovskaya, E. L.",
    title = "{Parton transport and hadronization from the dynamical quasiparticle point of view}",
    eprint = "0808.0022",
    archivePrefix = "arXiv",
    primaryClass = "hep-ph",
    doi = "10.1103/PhysRevC.78.034919",
    journal = "Phys. Rev. C",
    volume = "78",
    pages = "034919",
    year = "2008"
}

@article{Bratkovskaya:2011wp,
    author = "Bratkovskaya, E. L. and Cassing, W. and Konchakovski, V. P. and Linnyk, O.",
    title = "{Parton-Hadron-String Dynamics at Relativistic Collider Energies}",
    eprint = "1101.5793",
    archivePrefix = "arXiv",
    primaryClass = "nucl-th",
    doi = "10.1016/j.nuclphysa.2011.03.003",
    journal = "Nucl. Phys. A",
    volume = "856",
    pages = "162",
    year = "2011"
}

@article{Linnyk:2015rco,
    author = "Linnyk, O. and Bratkovskaya, E. L. and Cassing, W.",
    title = "{Effective QCD and transport description of dilepton and photon production in heavy-ion collisions and elementary processes}",
    eprint = "1512.08126",
    archivePrefix = "arXiv",
    primaryClass = "nucl-th",
    doi = "10.1016/j.ppnp.2015.12.003",
    journal = "Prog. Part. Nucl. Phys.",
    volume = "87",
    pages = "50",
    year = "2016"
}

@article{Song:2018xca,
    author = "Song, Taesoo and Cassing, Wolfgang and Moreau, Pierre and Bratkovskaya, Elena",
    title = "{Open charm and dileptons from relativistic heavy-ion collisions}",
    eprint = "1803.02698",
    archivePrefix = "arXiv",
    primaryClass = "nucl-th",
    doi = "10.1103/PhysRevC.97.064907",
    journal = "Phys. Rev. C",
    volume = "97",
    pages = "064907",
    year = "2018"
}

@article{Okorokov:2017pxd,
    author = "Okorokov, V. A.",
    title = "{Chiral Effects in Nucleus --- Nucleus Collisions: Experimental Review}",
    doi = "10.1134/S1063778817060163",
    journal = "Phys. Atom. Nucl.",
    volume = "80",
    number = "6",
    pages = "1133",
    year = "2017"
}

@article{Landsteiner:2016led,
    author = "Landsteiner, Karl",
    title = "{Notes on Anomaly Induced Transport}",
    eprint = "1610.04413",
    archivePrefix = "arXiv",
    primaryClass = "hep-th",
    reportNumber = "IFT-UAM-CSIC-16-103",
    doi = "10.5506/APhysPolB.47.2617",
    journal = "Acta Phys. Polon. B",
    volume = "47",
    pages = "2617",
    year = "2016"
}

@article{Kharzeev:2015znc,
    author = "Kharzeev, D. E. and Liao, J. and Voloshin, S. A. and Wang, G.",
    title = "{Chiral magnetic and vortical effects in high-energy nuclear collisions: A status report}",
    eprint = "1511.04050",
    archivePrefix = "arXiv",
    primaryClass = "hep-ph",
    doi = "10.1016/j.ppnp.2016.01.001",
    journal = "Prog. Part. Nucl. Phys.",
    volume = "88",
    pages = "1",
    year = "2016"
}

@article{FCC:2018bvk,
    author = "Abada, A. and others",
    collaboration = "FCC",
    title = "{HE-LHC: The High-Energy Large Hadron Collider}: {Future Circular Collider Conceptual Design Report Volume 4}",
    reportNumber = "CERN-ACC-2018-0059",
    doi = "10.1140/epjst/e2019-900088-6",
    journal = "Eur. Phys. J. ST",
    volume = "228",
    number = "5",
    pages = "1109",
    year = "2019"
}

@article{FCC:2018vvp,
    author = "Abada, A. and others",
    collaboration = "FCC",
    title = "{FCC-hh: The Hadron Collider}: {Future Circular Collider Conceptual Design Report Volume 3}",
    reportNumber = "CERN-ACC-2018-0058",
    doi = "10.1140/epjst/e2019-900087-0",
    journal = "Eur. Phys. J. ST",
    volume = "228",
    number = "4",
    pages = "755",
    year = "2019"
}

@article{Okorokov:2019bsi,
    author = "Okorokov, V. A.",
    title = "{Magnetic field in nuclear collisions at ultra high energies}",
    eprint = "1906.00383",
    archivePrefix = "arXiv",
    primaryClass = "nucl-th",
    doi = "10.3390/physics1020017",
    journal = "MDPI Physics",
    volume = "1",
    number = "2",
    pages = "183",
    year = "2019"
}

@article{Zhao:2014aja,
    author = "Zhao, Feng",
    collaboration = "STAR",
    title = "{$\Lambda(K_S^0)$–h$^\pm$ and $\Lambda$-p azimuthal correlations with respect to event plane and search for chiral magnetic and vortical effects}",
    doi = "10.1016/j.nuclphysa.2014.08.108",
    journal = "Nucl. Phys. A",
    volume = "931",
    pages = "746",
    year = "2014"
}

@article{Lisa:2021zkj,
    author = "Lisa, Michael Annan and Barbon, Jo\~ao Guilherme Prado and Chinellato, David Dobrigkeit and Serenone, Willian Matioli and Shen, Chun and Takahashi, Jun and Torrieri, Giorgio",
    title = "{Vortex rings from high energy central p+A collisions}",
    eprint = "2101.10872",
    archivePrefix = "arXiv",
    primaryClass = "hep-ph",
    doi = "10.1103/PhysRevC.104.L011901",
    journal = "Phys. Rev. C",
    volume = "104",
    number = "1",
    pages = "011901",
    year = "2021"
}

@article{Buzzegoli:2021jeh,
    author = "Buzzegoli, M. and Kharzeev, Dmitri E.",
    title = "{Anomalous gravitomagnetic moment and nonuniversality of the axial vortical effect at finite temperature}",
    eprint = "2102.01676",
    archivePrefix = "arXiv",
    primaryClass = "hep-th",
    doi = "10.1103/PhysRevD.103.116005",
    journal = "Phys. Rev. D",
    volume = "103",
    number = "11",
    pages = "116005",
    year = "2021"
}

@article{Greisen:1966jv,
    author = "Greisen, Kenneth",
    title = "{End to the cosmic ray spectrum?}",
    doi = "10.1103/PhysRevLett.16.748",
    journal = "Phys. Rev. Lett.",
    volume = "16",
    pages = "748",
    year = "1966"
}

@article{Zatsepin:1966jv,
    author = "Zatsepin, G. T. and Kuzmin, V. A.",
    title = "{Upper limit of the spectrum of cosmic rays}",
    journal = "JETP Lett.",
    volume = "4",
    pages = "78",
    year = "1966"
}

@article{RP-333-309-2000,
    author = "Watson, A. A.",
    title = "Ultra-high-energy cosmic rays: the experimental situation",
    doi = "10.1016/S0370-1573(00)00027-2",
    journal = "Phys. Rep.",
    volume = "333",
    pages = "309",
    year = "2000"
}

@article{PU-56-304-2013,
    author = "Troitskii, S. V.",
    title = "Cosmic particles with energies above $10^{19}$ eV: a brief summary of results",
    doi = "10.3367/UFNe.0183.201303i.0323",
    journal = "Phys. Usp.",
    volume = "56",
    pages = "304",
    year = "2013"
}

@article{Okorokov-PAN-82-838-2019,
    author = "Okorokov, V. A.",
    title = "Global characteristics of the medium produced in ultra-high energy cosmic ray collisions",
    eprint = "1907.02573",
    archivePrefix = "arXiv",
    primaryClass = "astro-ph.HE",
    doi = "10.1134/S1063778819660426",
    journal = "Phys. At. Nucl.",
    volume = "82",
    pages = "838",
    year = "2019"
}

@article{Kalaydzhyan:2014uha,
    author = "Kalaydzhyan, Tigran and Shuryak, Edward",
    title = "{''Explosive regime'' should dominate collisions of ultra-high energy cosmic rays}",
    eprint = "1407.3270",
    archivePrefix = "arXiv",
    primaryClass = "hep-ph",
    month = "7",
    year = "2014"
}

@article{EAS-MSU:2019kmv,
    author = "Dembinski, H. P. and others",
    editor = "Lhenry-Yvon, I. and Biteau, J. and Biteau, O. and Ghia, P.",
    collaboration = "EAS-MSU, IceCube, KASCADE-Grande, NEVOD-DECOR, Pierre Auger, SUGAR, Telescope Array, Yakutsk EAS Array",
    title = "{Report on Tests and Measurements of Hadronic Interaction Properties with Air Showers}",
    eprint = "1902.08124",
    archivePrefix = "arXiv",
    primaryClass = "astro-ph.HE",
    doi = "10.1051/epjconf/201921002004",
    journal = "EPJ Web Conf.",
    volume = "210",
    pages = "02004",
    year = "2019"
}

@article{Pierog:2020ghc,
    author = "Pierog, Tanguy and Baur, Sebastian and Dembinski, Hans P. and Ulrich, Ralf and Werner, Klaus",
    title = "{Collective Hadronization and Air Showers: Can LHC Data Solve the Muon Puzzle ?}",
    doi = "10.22323/1.358.0387",
    journal = "PoS",
    volume = "ICRC2019",
    pages = "387",
    year = "2020"
}

@article{arxiv-2006.11110-2020,
    author = "Sirunyan, A. M. and other",
    collaboration = "CMS",
    title = "Evidence for top quark production in nucleus-nucleus collisions",
    eprint = "2006.11110",
    archivePrefix = "arXiv",
    primaryClass = "hep-ph",
    doi = "10.1103/PhysRevLett.125.222001",
    journal = "Phys. Rev. Lett.",
    volume = "125",
    pages = "222001",
    year = "2020"
}

@article{Okorokov-JPCS-1690-012006-2020,
    author = "Okorokov, V. A.",
    title = "Top pair production at ultra-high energies",
    eprint = "2010.03912",
    archivePrefix = "arXiv",
    primaryClass = "hep-ph",
    doi = "10.1088/1742-6596/1690/1/012006",
    journal = "J. Phys. Conf. Ser.",
    volume = "690",
    pages = "012006",
    year = "2020"
}

@article{dEnterria:2018kcz,
    author = "d'Enterria, David and Pierog, Tanguy and Sun, Guanhao",
    title = "{Impact of QCD jets and heavy-quark production in cosmic-ray proton atmospheric showers up to 10$^{20}$ eV}",
    eprint = "1809.06406",
    archivePrefix = "arXiv",
    primaryClass = "astro-ph.HE",
    doi = "10.3847/1538-4357/ab01e2",
    journal = "Astrophys. J.",
    volume = "874",
    pages = "152",
    year = "2019"
}

@article{Abdallah:2003xd,
      author         = "{DELPHI Collaboration}",
      title          = "{Study of tau-pair production in photon-photon collisions
                        at LEP and limits on the anomalous electromagnetic moments
                        of the tau lepton}",
      collaboration  = "DELPHI",
      journal        = "Eur. Phys. J.",
      volume         = "C35",
      year           = "2004",
      pages          = "159",
      doi            = "10.1140/epjc/s2004-01852-y",
      eprint         = "hep-ex/0406010",
      archivePrefix  = "arXiv",
      primaryClass   = "hep-ex",
      reportNumber   = "CERN-EP-2003-058",
      SLACcitation   = "%%CITATION = HEP-EX/0406010;%%"
}

@article{Dyndal:2020yen,
    author = "Dyndal, Mateusz and Klusek-Gawenda, Mariola and Schott, Matthias and Szczurek, Antoni",
    title = "{Anomalous electromagnetic moments of $\tau$ lepton in $\gamma \gamma \to \tau^+ \tau^-$ reaction in Pb+Pb collisions at the LHC}",
    eprint = "2002.05503",
    archivePrefix = "arXiv",
    primaryClass = "hep-ph",
    doi = "10.1016/j.physletb.2020.135682",
    journal = "Phys. Lett. B",
    volume = "809",
    pages = "135682",
    year = "2020"
}

@article{Beresford:2019gww,
    author = "Beresford, Lydia and Liu, Jesse",
    title = "{New physics and tau $g-2$ using LHC heavy ion collisions}",
    eprint = "1908.05180",
    archivePrefix = "arXiv",
    primaryClass = "hep-ph",
    doi = "10.1103/PhysRevD.102.113008",
    journal = "Phys. Rev. D",
    volume = "102",
    pages = "113008",
    year = "2020"
}

@article{Martin:2001st,
      author         = "Martin, Stephen P. and Wells, James D.",
      title          = "{Muon Anomalous Magnetic Dipole Moment in Supersymmetric
                        Theories}",
      journal        = "Phys. Rev.",
      volume         = "D64",
      year           = "2001",
      pages          = "035003",
      doi            = "10.1103/PhysRevD.64.035003",
      eprint         = "hep-ph/0103067",
      archivePrefix  = "arXiv",
      primaryClass   = "hep-ph",
      reportNumber   = "FERMILAB-PUB-01-030-T, LBNL-47586",
      SLACcitation   = "%%CITATION = HEP-PH/0103067;%%"
}

@article{Feruglio:2018fxo,
      author         = "Feruglio, Ferruccio and Paradisi, Paride and Sumensari, Olcyr",
      title          = "{Implications of scalar and tensor explanations of $R_{D^{(\ast)}}$}",
      journal        = "JHEP",
      volume         = "11",
      year           = "2018",
      pages          = "191",
      doi            = "10.1007/JHEP11(2018)191",
      eprint         = "1806.10155",
      archivePrefix  = "arXiv",
      primaryClass   = "hep-ph",
      SLACcitation   = "%%CITATION = ARXIV:1806.10155;%%"
}

@article{Moyotl:2012zz,
      author         = "Moyotl, A. and Tavares-Velasco, G.",
      title          = "{Weak properties of the tau lepton via a spin-0                       unparticle}",
      journal        = "Phys. Rev.",
      volume         = "D86",
      year           = "2012",
      pages          = "013014",
      doi            = "10.1103/PhysRevD.86.013014",
      eprint         = "1210.1994",
      archivePrefix  = "arXiv",
      primaryClass   = "hep-ph",
      SLACcitation   = "%%CITATION = ARXIV:1210.1994;%%"
}

@article{GutierrezRodriguez:2004ch,
      author         = "Gutierrez-Rodriguez, A. and Hernandez-Ruiz, M. A. and Luis-Noriega, L. N.",
      title          = "{Limits on the dipole moments of the tau lepton via the process $e^+ e^- \to \tau^+ \tau^- \gamma$ in a left right symmetric model}",
      journal        = "Mod. Phys. Lett.",
      volume         = "A19",
      year           = "2004",
      pages          = "2227",
      doi            = "10.1142/S0217732304014689",
      eprint         = "hep-ph/0403237",
      archivePrefix  = "arXiv",
      primaryClass   = "hep-ph",
      SLACcitation   = "%%CITATION = HEP-PH/0403237;%%"
}

@inproceedings{Burmasov:2022gnl,
    author = "Burmasov, Nazar and Kryshen, Evgeny and Buehler, Paul and Lavicka, Roman",
    title = "{Feasibility of tau $g-2$ measurements in ultra-peripheral collisions of heavy ions}",
    eprint = "2203.00990",
    archivePrefix = "arXiv",
    primaryClass = "hep-ph",
    month = "3",
    year = "2022"
}

@inproceedings{dEnterria:2021ljz,
    author = "d'Enterria, David",
    title = "{Collider constraints on axion-like particles}",
    booktitle = "{Workshop on Feebly Interacting Particles}",
    eprint = "2102.08971",
    archivePrefix = "arXiv",
    primaryClass = "hep-ex",
    month = "2",
    year = "2021"
}

@article{Bauer:2017ris,
    author = "Bauer, Martin and Neubert, Matthias and Thamm, Andrea",
    title = "{Collider Probes of Axion-Like Particles}",
    eprint = "1708.00443",
    archivePrefix = "arXiv",
    primaryClass = "hep-ph",
    reportNumber = "MITP-17-047",
    doi = "10.1007/JHEP12(2017)044",
    journal = "JHEP",
    volume = "12",
    pages = "044",
    year = "2017"
}

@article{CMS:2018erd,
    author = "Sirunyan, Albert M and others",
    collaboration = "CMS",
    title = "{Evidence for light-by-light scattering and searches for axion-like particles in ultraperipheral PbPb collisions at $\sqrt{s_\mathrm{NN}} =$ 5.02 TeV}",
    eprint = "1810.04602",
    archivePrefix = "arXiv",
    primaryClass = "hep-ex",
    reportNumber = "CMS-FSQ-16-012, CERN-EP-2018-271",
    doi = "10.1016/j.physletb.2019.134826",
    journal = "Phys. Lett. B",
    volume = "797",
    pages = "134826",
    year = "2019"
}

@article{ATLAS:2020hii,
    author = "Aad, Georges and others",
    collaboration = "ATLAS",
    title = "{Measurement of light-by-light scattering and search for axion-like particles with 2.2 nb$^{-1}$ of Pb+Pb data with the ATLAS detector}",
    eprint = "2008.05355",
    archivePrefix = "arXiv",
    primaryClass = "hep-ex",
    reportNumber = "CERN-EP-2020-135",
    doi = "10.1007/JHEP03(2021)243",
    journal = "JHEP",
    volume = "03",
    pages = "243",
    year = "2021"
}

@article{Marciano:2016yhf,
    author = "Marciano, W. J. and Masiero, A. and Paradisi, P. and Passera, M.",
    title = "{Contributions of axionlike particles to lepton dipole moments}",
    eprint = "1607.01022",
    archivePrefix = "arXiv",
    primaryClass = "hep-ph",
    doi = "10.1103/PhysRevD.94.115033",
    journal = "Phys. Rev. D",
    volume = "94",
    pages = "115033",
    year = "2016"
}

@article{Klusek-Gawenda:2013rtu,
    author = "Klusek-Gawenda, Mariola and Szczurek, Antoni",
    title = "{$\pi^+ \pi^-$ and $\pi^0 \pi^0$ pair production in photon-photon and in ultraperipheral ultrarelativistic heavy ion collisions}",
    eprint = "1302.4204",
    archivePrefix = "arXiv",
    primaryClass = "nucl-th",
    doi = "10.1103/PhysRevC.87.054908",
    journal = "Phys. Rev. C",
    volume = "87",
    pages = "054908",
    year = "2013"
}

@article{Goncalves:2021pdc,
    author = "Goncalves, V. P. and Martins, D. E. and Rangel, M. S.",
    title = "{Searching for axionlike particles with low masses in pPb and PbPb collisions}",
    eprint = "2103.01862",
    archivePrefix = "arXiv",
    primaryClass = "hep-ph",
    doi = "10.1140/epjc/s10052-021-09314-2",
    journal = "Eur. Phys. J. C",
    volume = "81",
    pages = "522",
    year = "2021"
}

@article{Ho:2021uem,
    author = "Ho, David L.-J. and Rajantie, Arttu",
    title = "{Instanton solution for Schwinger production of \textquoteright{}t Hooft-Polyakov monopoles}",
    eprint = "2103.12799",
    archivePrefix = "arXiv",
    primaryClass = "hep-th",
    reportNumber = "IMPERIAL-TP-2021-DH-04",
    doi = "10.1103/PhysRevD.103.115033",
    journal = "Phys. Rev. D",
    volume = "103",
    pages = "115033",
    year = "2021"
}

@article{Adamova:2019vkf,
    author = "Adamov\'a, D. and others",
    title = "{A next-generation LHC heavy-ion experiment}",
    eprint = "1902.01211",
    archivePrefix = "arXiv",
    primaryClass = "physics.ins-det",
    month = "1",
    year = "2019"
}

@article{Schwinger:1951nm,
    author = "Schwinger, Julian S.",
    editor = "Milton, K. A.",
    title = "{On gauge invariance and vacuum polarization}",
    doi = "10.1103/PhysRev.82.664",
    journal = "Phys. Rev.",
    volume = "82",
    pages = "664",
    year = "1951"
}

@article{Affleck:1981ag,
    author = "Affleck, Ian K. and Manton, Nicholas S.",
    title = "{Monopole Pair Production in a Magnetic Field}",
    reportNumber = "MIT-CTP-933",
    doi = "10.1016/0550-3213(82)90511-9",
    journal = "Nucl. Phys. B",
    volume = "194",
    pages = "38",
    year = "1982"
}

@article{Dirac:1931kp,
    author = "Dirac, Paul Adrien Maurice",
    title = "{Quantised singularities in the electromagnetic field,}",
    reportNumber = "RX-722",
    doi = "10.1098/rspa.1931.0130",
    journal = "Proc. Roy. Soc. Lond. A",
    volume = "133",
    pages = "60",
    year = "1931"
}

@article{Huang:2016,
	doi = {10.1088/0034-4885/79/7/076302},
	year = 2016,
	publisher = {{IOP} Publishing},
	volume = {79},
	pages = {076302},
	author = {Xu-Guang Huang},
	title = {Electromagnetic fields and anomalous transports in heavy-ion collisions{\textemdash}a pedagogical review},
	journal = {Reports on Progress in Physics}
}

@article{Witten:1979kh,
    author = "Witten, Edward",
    title = "{Baryons in the 1/n Expansion}",
    reportNumber = "HUTP-79-A007",
    doi = "10.1016/0550-3213(79)90232-3",
    journal = "Nucl. Phys. B",
    volume = "160",
    pages = "57",
    year = "1979"
}

@article{Drukier:1982,
  title = {Monopole Pair Creation in Energetic Collisions: Is It Possible?},
  author = {Drukier, A. K. and Nussinov, S.},
  journal = {Phys. Rev. Lett.},
  volume = {49},
  issue = {2},
  pages = {102},
  year = {1982},
  publisher = {American Physical Society},
  doi = {10.1103/PhysRevLett.49.102},
}

@article{Milton:2006cp,
    author = "Milton, Kimball A.",
    title = "{Theoretical and experimental status of magnetic monopoles}",
    eprint = "hep-ex/0602040",
    archivePrefix = "arXiv",
    doi = "10.1088/0034-4885/69/6/R02",
    journal = "Rept. Prog. Phys.",
    volume = "69",
    pages = "1637",
    year = "2006"
}

@article{MoEDAL:2014ttp,
    author = "Acharya, B. and others",
    collaboration = "MoEDAL",
    title = "{The Physics Programme Of The MoEDAL Experiment At The LHC}",
    eprint = "1405.7662",
    archivePrefix = "arXiv",
    primaryClass = "hep-ph",
    reportNumber = "KCL-PH-TH-2014-02, LCTS-2014-02, CERN-PH-TH-2014-021, IFIC-14-16, IMPERIAL-TP-2014-AR-1, KCL-PH-TH/2014-02, LCTS/2014-02, CERN-PH-TH/2014-021, IFIC/14-16,
  Imperial/TP/2014/AR/1",
    doi = "10.1142/S0217751X14300506",
    journal = "Int. J. Mod. Phys. A",
    volume = "29",
    pages = "1430050",
    year = "2014"
}

@article{MoEDAL:2021vix,
    author = "Acharya, B. and others",
    collaboration = "MoEDAL",
    title = "{Search for magnetic monopoles produced via the Schwinger mechanism}",
    eprint = "2106.11933",
    archivePrefix = "arXiv",
    primaryClass = "hep-ex",
    doi = "10.1038/s41586-021-04298-1",
    journal = "Nature",
    volume = "602",
    number = "7895",
    pages = "63",
    year = "2022"
}

@article{Gould:2021bre,
    author = "Gould, Oliver and Ho, David L.-J. and Rajantie, Arttu",
    title = "{Schwinger pair production of magnetic monopoles: Momentum distribution for heavy-ion collisions}",
    eprint = "2103.14454",
    archivePrefix = "arXiv",
    primaryClass = "hep-ph",
    reportNumber = "HIP-2021-12/TH, IMPERIAL-TP-2021-DH-05",
    doi = "10.1103/PhysRevD.104.015033",
    journal = "Phys. Rev. D",
    volume = "104",
    pages = "015033",
    year = "2021"
}

@article{King:2016pys,
    author = "King, Matthew",
    editor = "Aguilar-Ben\'\i{}tez, M and Fuster, J and Mart\'\i{}-Garc\'\i{}a, S and Santamar\'\i{}a, A",
    collaboration = "MoEDAL",
    title = "{Simulation of the MoEDAL experiment}",
    doi = "10.1016/j.nuclphysbps.2015.09.459",
    journal = "Nucl. Part. Phys. Proc.",
    volume = "273-275",
    pages = "2560",
    year = "2016"
}

@article{Clemencic:2011zza,
    author = "Clemencic, M. and Corti, G. and Easo, S. and Jones, C. R. and Miglioranzi, S. and Pappagallo, M. and Robbe, P.",
    editor = "Lin, Simon C.",
    collaboration = "LHCb",
    title = "{The LHCb simulation application, Gauss: Design, evolution and experience}",
    doi = "10.1088/1742-6596/331/3/032023",
    journal = "J. Phys. Conf. Ser.",
    volume = "331",
    pages = "032023",
    year = "2011"
}

@article{EuropeanStrategyforParticlePhysicsPreparatoryGroup:2019qin,
    author = "Ellis, Richard Keith and others",
    title = "{Physics Briefing Book}: {Input for the European Strategy for Particle Physics Update 2020}",
    eprint = "1910.11775",
    archivePrefix = "arXiv",
    primaryClass = "hep-ex",
    reportNumber = "CERN-ESU-004",
    month = "10",
    year = "2019"
}

@article{Horvat:2020ycy,
    author = "Horvat, Raul and Latas, Duvsko and Trampetic, Josip and You, Jiangyang",
    title = "{Light-by-Light Scattering and Spacetime Noncommutativity}",
    eprint = "2002.01829",
    archivePrefix = "arXiv",
    primaryClass = "hep-ph",
    doi = "10.1103/PhysRevD.101.095035",
    journal = "Phys. Rev. D",
    volume = "101",
    pages = "095035",
    year = "2020"
}

@inbook{CidVidal:2018eel,
    author = "Cid Vidal, Xabier and others",
    editor = "Dainese, Andrea and Mangano, Michelangelo and Meyer, Andreas B. and Nisati, Aleandro and Salam, Gavin and Vesterinen, Mika Anton",
    title = "{Report from Working Group 3}: {Beyond the Standard Model physics at the HL-LHC and HE-LHC}",
    booktitle = "{Report on the Physics at the HL-LHC,and Perspectives for the HE-LHC}",
    eprint = "1812.07831",
    archivePrefix = "arXiv",
    primaryClass = "hep-ph",
    reportNumber = "CERN-LPCC-2018-05",
    doi = "10.23731/CYRM-2019-007.585",
    volume = "7",
    pages = "585",
    month = "12",
    year = "2019"
}

@article{He:1997pj,
      author         = "He, Y. D.",
      title          = "{Search for a Dirac magnetic monopole in high-energy
                        nucleus-nucleus collisions}",
      journal        = "Phys. Rev. Lett.",
      volume         = "79",
      year           = "1997",
      pages          = "3134",
      doi            = "10.1103/PhysRevLett.79.3134",
      SLACcitation   = "%%CITATION = PRLTA,79,3134;%%"
}

@article{Ellis:2017edi,
      author         = "Ellis, John and Mavromatos, Nick E. and You, Tevong",
      title          = "{Light-by-Light Scattering Constraint on Born-Infeld
                        Theory}",
      journal        = "Phys. Rev. Lett.",
      volume         = "118",
      year           = "2017",
      pages          = "261802",
      doi            = "10.1103/PhysRevLett.118.261802",
      eprint         = "1703.08450",
      archivePrefix  = "arXiv",
      primaryClass   = "hep-ph",
      reportNumber   = "CAVENDISH-HEP-17-04, DAMTP-2017-12, KCL-PH-TH-2017-11,
                        CERN-TH-2017-062",
      SLACcitation   = "%%CITATION = ARXIV:1703.08450;%%"
}

@article{Hewett:2000zp,
      author         = "Hewett, JoAnne L. and Petriello, Frank J. and Rizzo,
                        Thomas G.",
      title          = "{Signals for noncommutative interactions at linear
                        colliders}",
      journal        = "Phys. Rev.",
      volume         = "D64",
      year           = "2001",
      pages          = "075012",
      doi            = "10.1103/PhysRevD.64.075012",
      eprint         = "hep-ph/0010354",
      archivePrefix  = "arXiv",
      primaryClass   = "hep-ph",
      reportNumber   = "SLAC-PUB-8635, FERMILAB-PUB-00-286-T",
      SLACcitation   = "%%CITATION = HEP-PH/0010354;%%"
}

@article{Dvorkin:2019zdi,
      author         = "Dvorkin, Cora and Lin, Tongyan and Schutz, Katelin",
      title          = "{Making dark matter out of light: freeze-in from plasma
                        effects}",
      journal        = "Phys. Rev.",
      volume         = "D99",
      year           = "2019",
      pages          = "115009",
      doi            = "10.1103/PhysRevD.99.115009",
      eprint         = "1902.08623",
      archivePrefix  = "arXiv",
      primaryClass   = "hep-ph",
      SLACcitation   = "%%CITATION = ARXIV:1902.08623;%%"
}

@article{Farrar:2020zeo,
      author         = "Farrar, Glennys R. and Wang, Zihui and Xu, Xingchen",
      title          = "{Dark Matter Particle in QCD}",
      year           = "2020",
      eprint         = "2007.10378",
      archivePrefix  = "arXiv",
      primaryClass   = "hep-ph",
      SLACcitation   = "%%CITATION = ARXIV:2007.10378;%%"
}

@article{Drewes:2019vjy,
      author         = "Drewes, Marco and Giammanco, Andrea and Hajer, Jan and
                        Lucente, Michele",
      title          = "{New long-lived particle searches in heavy-ion collisions
                        at the LHC}",
      journal        = "Phys. Rev.",
      volume         = "D101",
      year           = "2020",
      pages          = "055002",
      doi            = "10.1103/PhysRevD.101.055002",
      eprint         = "1905.09828",
      archivePrefix  = "arXiv",
      primaryClass   = "hep-ph",
      reportNumber   = "CP3-19-26",
      SLACcitation   = "%%CITATION = ARXIV:1905.09828;%%"
}

@article{Drewes:2018xma,
      author         = "Drewes, Marco and Giammanco, Andrea and Hajer, Jan and
                        Lucente, Michele and Mattelaer, Olivier",
      title          = "{Searching for New Long Lived Particles in Heavy Ion
                        Collisions at the LHC}",
      journal        = "Phys. Rev. Lett.",
      volume         = "124",
      year           = "2020",
      pages          = "081801",
      doi            = "10.1103/PhysRevLett.124.081801",
      eprint         = "1810.09400",
      archivePrefix  = "arXiv",
      primaryClass   = "hep-ph",
      reportNumber   = "CP3-18-60",
      SLACcitation   = "%%CITATION = ARXIV:1810.09400;%%"
}

@article{Sirunyan:2019cgy,
      author         = "Sirunyan, Albert M and others",
      title          = "{Pseudorapidity distributions of charged hadrons in
                        xenon-xenon collisions at $\sqrt{s_\mathrm{NN}} =$ 5.44
                        TeV}",
      collaboration  = "CMS",
      journal        = "Phys. Lett.",
      volume         = "B799",
      year           = "2019",
      pages          = "135049",
      doi            = "10.1016/j.physletb.2019.135049",
      eprint         = "1902.03603",
      archivePrefix  = "arXiv",
      primaryClass   = "hep-ex",
      reportNumber   = "CMS-HIN-17-006, CERN-EP-2018-294",
      SLACcitation   = "%%CITATION = ARXIV:1902.03603;%%"
}

@article{Sirunyan:2017xku,
      author         = "Sirunyan, Albert M and others",
      title          = "{Observation of top quark production in proton-nucleus
                        collisions}",
      collaboration  = "CMS",
      journal        = "Phys. Rev. Lett.",
      volume         = "119",
      year           = "2017",
      pages          = "242001",
      doi            = "10.1103/PhysRevLett.119.242001",
      eprint         = "1709.07411",
      archivePrefix  = "arXiv",
      primaryClass   = "nucl-ex",
      reportNumber   = "CMS-HIN-17-002, CERN-EP-2017-239",
      SLACcitation   = "%%CITATION = ARXIV:1709.07411;%%"
}

@article{Gould:2019myj,
      author         = "Gould, Oliver and Ho, David L.-J. and Rajantie, Arttu",
      title          = "{Towards Schwinger production of magnetic monopoles in
                        heavy-ion collisions}",
      journal        = "Phys. Rev.",
      volume         = "D100",
      year           = "2019",
      pages          = "015041",
      doi            = "10.1103/PhysRevD.100.015041",
      eprint         = "1902.04388",
      archivePrefix  = "arXiv",
      primaryClass   = "hep-th",
      reportNumber   = "IMPERIAL-TP-2019-DH-01, HIP-2019-2/TH,
                        IMPERIAL-TP-2019-DH-01; HIP-2019-2/TH",
      SLACcitation   = "%%CITATION = ARXIV:1902.04388;%%"
}

@article{Adams:2005cu,
      author         = "Abelev, B. I. and others",
      title          = "{Strangelet search at RHIC}",
      collaboration  = "STAR",
      journal        = "Phys. Rev.",
      volume         = "C76",
      year           = "2007",
      pages          = "011901",
      doi            = "10.1103/PhysRevC.76.011901",
      eprint         = "nucl-ex/0511047",
      archivePrefix  = "arXiv",
      primaryClass   = "nucl-ex",
      SLACcitation   = "%%CITATION = NUCL-EX/0511047;%%"
}

@article{Gould:2017zwi,
      author         = "Gould, Oliver and Rajantie, Arttu",
      title          = "{Magnetic monopole mass bounds from heavy ion collisions
                        and neutron stars}",
      journal        = "Phys. Rev. Lett.",
      volume         = "119",
      year           = "2017",
      pages          = "241601",
      doi            = "10.1103/PhysRevLett.119.241601",
      eprint         = "1705.07052",
      archivePrefix  = "arXiv",
      primaryClass   = "hep-ph",
      reportNumber   = "IMPERIAL-TP-2017-OG-2",
      SLACcitation   = "%%CITATION = ARXIV:1705.07052;%%"
}

@article{Kharzeev:2007jp,
      author         = "Kharzeev, Dmitri E. and McLerran, Larry D. and Warringa,
                        Harmen J.",
      title          = "{The Effects of topological charge change in heavy ion
                        collisions: Event by event P and CP violation}",
      journal        = "Nucl. Phys.",
      volume         = "A803",
      year           = "2008",
      pages          = "227-253",
      doi            = "10.1016/j.nuclphysa.2008.02.298",
      eprint         = "0711.0950",
      archivePrefix  = "arXiv",
      primaryClass   = "hep-ph",
      SLACcitation   = "%%CITATION = ARXIV:0711.0950;%%"
}

@article{Fukushima:2008xe,
    author = "Fukushima, Kenji and Kharzeev, Dmitri E. and Warringa, Harmen J.",
    title = "{The Chiral Magnetic Effect}",
    eprint = "0808.3382",
    archivePrefix = "arXiv",
    primaryClass = "hep-ph",
    doi = "10.1103/PhysRevD.78.074033",
    journal = "Phys. Rev. D",
    volume = "78",
    pages = "074033",
    year = "2008"
}

@article{Ho:2020ltr,
    author = "Ho, David L.-J. and Rajantie, Arttu",
    title = "{Electroweak sphaleron in a strong magnetic field}",
    eprint = "2005.03125",
    archivePrefix = "arXiv",
    primaryClass = "hep-th",
    reportNumber = "IMPERIAL-TP-2020-DH-01",
    doi = "10.1103/PhysRevD.102.053002",
    journal = "Phys. Rev. D",
    volume = "102",
    pages = "053002",
    year = "2020"
}

@article{Ho:2019ads,
    author = "Ho, David L.-J. and Rajantie, Arttu",
    title = "{Classical production of \textquoteright{}t Hooft\textendash{}Polyakov monopoles from magnetic fields}",
    eprint = "1911.06088",
    archivePrefix = "arXiv",
    primaryClass = "hep-th",
    reportNumber = "IMPERIAL-TP-2019-DH-02",
    doi = "10.1103/PhysRevD.101.055003",
    journal = "Phys. Rev. D",
    volume = "101",
    pages = "055003",
    year = "2020"
}

@article{Baltz:2007kq,
    author = "Baltz, A.J. and others",
    title = "{The Physics of Ultraperipheral Collisions at the LHC}",
    eprint = "0706.3356",
    archivePrefix = "arXiv",
    primaryClass = "nucl-ex",
    doi = "10.1016/j.physrep.2007.12.001",
    journal = "Phys. Rept.",
    volume = "458",
    pages = "1",
    year = "2008"
}

@article{Klein:2020fmr,
    author = "Klein, Spencer and Steinberg, Peter",
    title = "{Photonuclear and Two-photon Interactions at High-Energy Nuclear Colliders}",
    eprint = "2005.01872",
    archivePrefix = "arXiv",
    primaryClass = "nucl-ex",
    doi = "10.1146/annurev-nucl-030320-033923",
    journal = "Ann. Rev. Nucl. Part. Sci.",
    volume = "70",
    pages = "323",
    year = "2020"
}

@article{dEnterria:2013zqi,
    author = "d'Enterria, David and da Silveira, Gustavo G.",
    title = "{Observing light-by-light scattering at the Large Hadron Collider}",
    eprint = "1305.7142",
    archivePrefix = "arXiv",
    primaryClass = "hep-ph",
    doi = "10.1103/PhysRevLett.111.080405",
    journal = "Phys. Rev. Lett.",
    volume = "111",
    pages = "080405",
    year = "2013",
    relatedtype = "erratum",
    related = "dEnterria:2013zqi-1"
}

@article{Aad:2019ock,
    author = "Aad, Georges and others",
    collaboration = "ATLAS",
    title = "{Observation of light-by-light scattering in ultraperipheral Pb+Pb collisions with the ATLAS detector}",
    eprint = "1904.03536",
    archivePrefix = "arXiv",
    primaryClass = "hep-ex",
    reportNumber = "CERN-EP-2019-051",
    doi = "10.1103/PhysRevLett.123.052001",
    journal = "Phys. Rev. Lett.",
    volume = "123",
    pages = "052001",
    year = "2019"
}

@article{Knapen:2016moh,
    author = "Knapen, Simon and Lin, Tongyan and Lou, Hou Keong and Melia, Tom",
    title = "{Searching for Axionlike Particles with Ultraperipheral Heavy-Ion Collisions}",
    eprint = "1607.06083",
    archivePrefix = "arXiv",
    primaryClass = "hep-ph",
    doi = "10.1103/PhysRevLett.118.171801",
    journal = "Phys. Rev. Lett.",
    volume = "118",
    pages = "171801",
    year = "2017"
}

@article{Sirunyan:2018fhl,
    author = "Sirunyan, Albert M and others",
    collaboration = "CMS",
    title = "{Evidence for light-by-light scattering and searches for axion-like particles in ultraperipheral PbPb collisions at $\sqrt{s_\mathrm{NN}} =$ 5.02 TeV}",
    eprint = "1810.04602",
    archivePrefix = "arXiv",
    primaryClass = "hep-ex",
    reportNumber = "CMS-FSQ-16-012, CERN-EP-2018-271",
    doi = "10.1016/j.physletb.2019.134826",
    journal = "Phys. Lett. B",
    volume = "797",
    pages = "134826",
    year = "2019"
}

@article{Aaboud:2017bwk,
    author = "Aaboud, Morad and others",
    collaboration = "ATLAS",
    title = "{Evidence for light-by-light scattering in heavy-ion collisions with the ATLAS detector at the LHC}",
    eprint = "1702.01625",
    archivePrefix = "arXiv",
    primaryClass = "hep-ex",
    reportNumber = "CERN-EP-2016-316",
    doi = "10.1038/nphys4208",
    journal = "Nature Phys.",
    volume = "13",
    pages = "852",
    year = "2017"
}

@article{Goncalves:2020czp,
    author = "Goncalves, V.P. and Moreira, B.D.",
    title = "{Dark photons from pions produced in ultraperipheral $PbPb$ collisions}",
    eprint = "2006.08348",
    archivePrefix = "arXiv",
    primaryClass = "hep-ph",
    doi = "10.1016/j.physletb.2020.135635",
    journal = "Phys. Lett. B",
    volume = "808",
    pages = "135635",
    year = "2020"
}

@techreport{ATLAS:2020qfn,
    collaboration = "ATLAS",
    title = "{Observation of photon-induced $W^+W^-$ production in $pp$ collisions at $\sqrt{s}=13$ TeV using the ATLAS detector}",
    reportNumber = "ATLAS-CONF-2020-038",
    month = "8",
    year = "2020",
    url = "https://cds.cern.ch/record/2727859",
}

@article{dEnterria:2019jty,
    author = "d'Enterria, David and Martins, Daniel E. and Rebello Teles, Patricia",
    title = "{Higgs boson production in photon-photon interactions with proton, light-ion, and heavy-ion beams at current and future colliders}",
    eprint = "1904.11936",
    archivePrefix = "arXiv",
    primaryClass = "hep-ph",
    doi = "10.1103/PhysRevD.101.033009",
    journal = "Phys. Rev. D",
    volume = "101",
    pages = "033009",
    year = "2020"
}

@article{Babu:2016rcr,
    author = "Babu, K.S. and Jana, Sudip",
    title = "{Probing Doubly Charged Higgs Bosons at the LHC through Photon Initiated Processes}",
    eprint = "1612.09224",
    archivePrefix = "arXiv",
    primaryClass = "hep-ph",
    reportNumber = "OSU-HEP-16-11",
    doi = "10.1103/PhysRevD.95.055020",
    journal = "Phys. Rev. D",
    volume = "95",
    pages = "055020",
    year = "2017"
}

@article{Beresford:2018pbt,
    author = "Beresford, Lydia and Liu, Jesse",
    title = "{Search Strategy for Sleptons and Dark Matter Using the LHC as a Photon Collider}",
    eprint = "1811.06465",
    archivePrefix = "arXiv",
    primaryClass = "hep-ph",
    doi = "10.1103/PhysRevLett.123.141801",
    journal = "Phys. Rev. Lett.",
    volume = "123",
    pages = "141801",
    year = "2019"
}

@article{Harland-Lang:2018hmi,
    author = "Harland-Lang, L.A. and Khoze, V.A. and Ryskin, M.G. and Tasevsky, M.",
    title = "{LHC Searches for Dark Matter in Compressed Mass Scenarios: Challenges in the Forward Proton Mode}",
    eprint = "1812.04886",
    archivePrefix = "arXiv",
    primaryClass = "hep-ph",
    reportNumber = "IPP/18/103",
    doi = "10.1007/JHEP04(2019)010",
    journal = "JHEP",
    volume = "04",
    pages = "010",
    year = "2019"
}

@article{Godunov:2019jib,
    author = "Godunov, S.I. and Novikov, V.A. and Rozanov, A.N. and Vysotsky, M.I. and Zhemchugov, E.V.",
    title = "{Quasistable charginos in ultraperipheral proton-proton collisions at the LHC}",
    eprint = "1906.08568",
    archivePrefix = "arXiv",
    primaryClass = "hep-ph",
    doi = "10.1007/JHEP01(2020)143",
    journal = "JHEP",
    volume = "01",
    pages = "143",
    year = "2020"
}

@article{ATLAS:2020bxl,
    collaboration = "ATLAS",
    title = "{Observation and measurement of forward proton scattering in association with lepton pairs produced via the photon fusion mechanism at ATLAS}",
    reportNumber = "ATLAS-CONF-2020-041",
    month = "8",
    year = "2020",
    url = "http://cds.cern.ch/record/2727863",
}

@article{Cms:2018het,
    author = "Sirunyan, Albert M and others",
    collaboration = "CMS, TOTEM",
    title = "{Observation of proton-tagged, central (semi)exclusive production of high-mass lepton pairs in pp collisions at 13 TeV with the CMS-TOTEM precision proton spectrometer}",
    eprint = "1803.04496",
    archivePrefix = "arXiv",
    primaryClass = "hep-ex",
    reportNumber = "CMS-PPS-17-001, TOTEM 2018-001, CERN-EP-2018-014",
    doi = "10.1007/JHEP07(2018)153",
    journal = "JHEP",
    volume = "07",
    pages = "153",
    year = "2018"
}

@article{Khachatryan:2016mud,
    author = "Khachatryan, Vardan and others",
    collaboration = "CMS",
    title = "{Evidence for exclusive $\gamma\gamma \to W^+ W^-$ production and constraints on anomalous quartic gauge couplings in $pp$ collisions at $ \sqrt{s}=7 $ and 8 TeV}",
    eprint = "1604.04464",
    archivePrefix = "arXiv",
    primaryClass = "hep-ex",
    reportNumber = "CMS-FSQ-13-008, CERN-EP-2016-073",
    doi = "10.1007/JHEP08(2016)119",
    journal = "JHEP",
    volume = "08",
    pages = "119",
    year = "2016"
}

@article{Gonderinger:2010yn,
    author = "Gonderinger, Matthew and Ramsey-Musolf, Michael J.",
    title = "{Electron-to-Tau Lepton Flavor Violation at the Electron-Ion Collider}",
    eprint = "1006.5063",
    archivePrefix = "arXiv",
    primaryClass = "hep-ph",
    reportNumber = "NPAC-10-09",
    doi = "10.1007/JHEP11(2010)045",
    journal = "JHEP",
    volume = "11",
    pages = "045",
    year = "2010",
    note = "[Erratum: JHEP 05, 047 (2012)]"
}

@article{Cirigliano:2021img,
    author = "Cirigliano, Vincenzo and Fuyuto, Kaori and Lee, Christopher and Mereghetti, Emanuele and Yan, Bin",
    title = "{Charged Lepton Flavor Violation at the EIC}",
    eprint = "2102.06176",
    archivePrefix = "arXiv",
    primaryClass = "hep-ph",
    reportNumber = "LA-UR-21-20531",
    doi = "10.1007/JHEP03(2021)256",
    journal = "JHEP",
    volume = "03",
    pages = "256",
    year = "2021"
}

@article{Davoudiasl:2021mjy,
    author = "Davoudiasl, Hooman and Marcarelli, Roman and Neil, Ethan T.",
    title = "{Lepton-Flavor-Violating ALPs at the Electron-Ion Collider: A Golden Opportunity}",
    eprint = "2112.04513",
    archivePrefix = "arXiv",
    primaryClass = "hep-ph",
    month = "12",
    year = "2021"
}

@article{Liu:2021lan,
    author = "Liu, Yandong and Yan, Bin",
    title = "{Searching for the axion-like particle at the EIC}",
    eprint = "2112.02477",
    archivePrefix = "arXiv",
    primaryClass = "hep-ph",
    reportNumber = "LA-UR-21-31766",
    month = "12",
    year = "2021"
}

@article{AbdulKhalek:2021gbh,
    author = "Abdul Khalek, R. and others",
    title = "{Science Requirements and Detector Concepts for the Electron-Ion Collider: EIC Yellow Report}",
    eprint = "2103.05419",
    archivePrefix = "arXiv",
    primaryClass = "physics.ins-det",
    reportNumber = "BNL-220990-2021-FORE, JLAB-PHY-21-3198, LA-UR-21-20953",
    month = "3",
    year = "2021"
}

@article{Acharya:2018ohw,
      author         = "Acharya, Shreyasi and others",
      title          = "{Dielectron production in proton-proton collisions at $
                        \sqrt{s}=7 $ TeV}",
      collaboration  = "ALICE",
      journal        = "JHEP",
      volume         = "09",
      year           = "2018",
      pages          = "064",
      doi            = "10.1007/JHEP09(2018)064",
      eprint         = "1805.04391",
      archivePrefix  = "arXiv",
      primaryClass   = "hep-ex",
      reportNumber   = "CERN-EP-2018-102",
      SLACcitation   = "%%CITATION = ARXIV:1805.04391;%%"
}

@article{dEnterria:2009cwl,
    author = "d'Enterria, David and Lansberg, Jean-Philippe",
    title = "{Study of Higgs boson production and its b anti-b decay in gamma-gamma processes in proton-nucleus collisions at the LHC}",
    eprint = "0909.3047",
    archivePrefix = "arXiv",
    primaryClass = "hep-ph",
    reportNumber = "SLAC-PUB-13786",
    doi = "10.1103/PhysRevD.81.014004",
    journal = "Phys. Rev. D",
    volume = "81",
    pages = "014004",
    year = "2010"
}

@article{Agrawal:2021dbo,
    author = "Agrawal, Prateek and others",
    title = "{Feebly-interacting particles: FIPs 2020 workshop report}",
    eprint = "2102.12143",
    archivePrefix = "arXiv",
    primaryClass = "hep-ph",
    doi = "10.1140/epjc/s10052-021-09703-7",
    journal = "Eur. Phys. J. C",
    volume = "81",
    pages = "1015",
    year = "2021"
}

@article{dEnterria:2011twh,
    author = "d'Enterria, David and Engel, Ralph and Pierog, Tanguy and Ostapchenko, Sergey and Werner, Klaus",
    title = "{Constraints from the first LHC data on hadronic event generators for ultra-high energy cosmic-ray physics}",
    eprint = "1101.5596",
    archivePrefix = "arXiv",
    primaryClass = "astro-ph.HE",
    doi = "10.1016/j.astropartphys.2011.05.002",
    journal = "Astropart. Phys.",
    volume = "35",
    pages = "98",
    year = "2011"
}

@article{Kampert:2012mx,
    author = "Kampert, Karl-Heinz and Unger, Michael",
    title = "{Measurements of the Cosmic Ray Composition with Air Shower Experiments}",
    eprint = "1201.0018",
    archivePrefix = "arXiv",
    primaryClass = "astro-ph.HE",
    doi = "10.1016/j.astropartphys.2012.02.004",
    journal = "Astropart. Phys.",
    volume = "35",
    pages = "660",
    year = "2012"
}

@article{dEnterria:2015mgr,
    author = "d'Enterria, David and Krajcz\'ar, Kriszti\'an and Paukkunen, Hannu",
    title = "{Top-quark production in proton\textendash{}nucleus and nucleus\textendash{}nucleus collisions at LHC energies and beyond}",
    eprint = "1501.05879",
    archivePrefix = "arXiv",
    primaryClass = "hep-ph",
    doi = "10.1016/j.physletb.2015.04.044",
    journal = "Phys. Lett. B",
    volume = "746",
    pages = "64",
    year = "2015"
}

@article{Albrecht:2021cxw,
    author = "Albrecht, Johannes and others",
    title = "{The Muon Puzzle in cosmic-ray induced air showers and its connection to the Large Hadron Collider}",
    eprint = "2105.06148",
    archivePrefix = "arXiv",
    primaryClass = "astro-ph.HE",
    month = "5",
    year = "2021"
}

@article{Schichtel:2019hfn,
    author = "Schichtel, Peter and Spannowsky, Michael and Waite, Philip",
    title = "{Constraining strongly coupled new physics from cosmic rays with machine learning techniques}",
    eprint = "1906.09064",
    archivePrefix = "arXiv",
    primaryClass = "hep-ph",
    reportNumber = "IPPP/19/39",
    doi = "10.1209/0295-5075/127/61002",
    journal = "EPL",
    volume = "127",
    number = "6",
    pages = "61002",
    year = "2019"
}

@article{Brooijmans:2016lfv,
    author = "Brooijmans, Gustaaf and Schichtel, Peter and Spannowsky, Michael",
    title = "{Cosmic ray air showers from sphalerons}",
    eprint = "1602.00647",
    archivePrefix = "arXiv",
    primaryClass = "hep-ph",
    reportNumber = "IPPP-16-09, DCPT-16-18, MCNET-16-02",
    doi = "10.1016/j.physletb.2016.08.030",
    journal = "Phys. Lett. B",
    volume = "761",
    pages = "213",
    year = "2016"
}

@article{Dainese:2016gch,
    author = "Dainese, A. and others",
    title = "{Heavy ions at the Future Circular Collider}",
    eprint = "1605.01389",
    archivePrefix = "arXiv",
    primaryClass = "hep-ph",
    reportNumber = "CERN-TH-2016-107",
    doi = "10.23731/CYRM-2017-003.635",
    month = "5",
    year = "2016"
}

@article{CMS:2021ncv,
    collaboration = "CMS",
    title = "{The CMS Precision Proton Spectrometer at the HL-LHC -- Expression of Interest}",
    eprint = "2103.02752",
    archivePrefix = "arXiv",
    primaryClass = "physics.ins-det",
    reportNumber = "CERN-CMS-NOTE-2020-008",
    month = "3",
    year = "2021"
}

@inproceedings{Irastorza:2021tdu,
    author = "Irastorza, Igor G.",
    title = "{An introduction to axions and their detection}",
    booktitle = "{Les Houches summer school on Dark Matter}",
    eprint = "2109.07376",
    archivePrefix = "arXiv",
    primaryClass = "hep-ph",
    month = "9",
    year = "2021"
}

@article{dEnterria:2016yhy,
    author = "d'Enterria, David and Snigirev, Alexander M.",
    title = "{Triple-parton scatterings in proton\textendash{}nucleus collisions at high energies}",
    eprint = "1612.08112",
    archivePrefix = "arXiv",
    primaryClass = "hep-ph",
    doi = "10.1140/epjc/s10052-018-5687-2",
    journal = "Eur. Phys. J. C",
    volume = "78",
    number = "5",
    pages = "359",
    year = "2018"
}

\end{document}